\documentclass{article} 

\usepackage[dvips]{epsfig,color}
\usepackage{graphicx}
\usepackage{amsmath,amsfonts,amsbsy,amssymb}

\begin{document}

	\begin{center}   
	{\bf  The Split-Algebras and Non-compact Hopf Maps\\}   
\vspace{0.3cm} 	
{\em Kazuki\ Hasebe\\}   
	\vspace{0.2cm}   
	{\small  Department of General Education,\\   
	Kagawa National College of Technology, \\   
        Takuma-cho, Mitoyo-city, Kagawa 769-1192, Japan \\}   
\vspace{0.2cm}      
     {\small email:  hasebe@dg.kagawa-nct.ac.jp\\} 

	\end{center}   

\begin{abstract} 
\noindent  

We develop a non-compact version of the Hopf maps based on the split-algebras. 
The split-algebras consist of three species: split-complex numbers, split-quaternions, and split-octonions.
They correspond to three non-compact Hopf maps that represent topological maps between hyperboloids in different dimensions with hyperboloid-bundle. 
We realize such non-compact Hopf maps in two ways: one is to utilize the split-imaginary unit, and the other is to utilize the ordinary imaginary unit.
Topological structures of the hyperboloid-bundles 
are explored, and the canonical connections are naturally regarded as non-compact gauge field of monopoles.   

\end{abstract}


\section{Introduction}\label{secintroduction}

In the sequel papers in 1931 and 1935 \cite{Hopf1931,Hopf1935}, Heinz Hopf introduced the notion of  topological maps from sphere to sphere in different dimensions: 
\begin{center}
\begin{tabular}{ccccccc}
\\ 
 & & $S^{3}$ &   $\overset{S^{1}}\longrightarrow $ & $S^{2}$ & & ~~~~~~~~~~(1st)\\
 &  $S^{7}$ & $\longrightarrow$ & $S^{4}$ &  & & ~~~~~~~~~~(2nd) \\
 $S^{15}$ & $\longrightarrow$ &   $S^{8}$ &&  & & ~~~~~~~~~~(3rd) \\  
\end{tabular}
\end{center}
Such maps are entitled as the Hopf maps in honor of his name. 
There are no other fibrations between spheres with sphere-bundle according to Adams' theorem \cite{Adams1960}. 
The importance of the Hopf maps is now widely appreciated in fibre-bundle theory of mathematics,
 and in applications of topology to physics \cite{Nakaharabook,TzeGurseyBook}. In particular, the three Hopf maps have direct relevance to monopoles; the 1st Hopf map is the underlying mathematics of Dirac's $U(1)$  monopole \cite{Dirac1931}, the 2nd Hopf map is that of Yang's $SU(2)$ monopole \cite{Yang1978}, and the 
3rd Hopf map is that of the $SO(8)$ monopole \cite{Grossman1984}. 
Interestingly, the Hopf maps are deeply related to the normed division algebras; $i.e.$ complex numbers $\mathbb{C}$, quaternions $\mathbb{H}$, and octonions $\mathbb{O}$ \cite{Baez2002}. 
In the language of the division algebras, the Hopf maps can be  
restated as 
\begin{center}
\begin{tabular}{ccccccc}
\\ 
 & & $S^1_{\mathbb{C}}$ &   $\overset{S^1_{\mathbb{R}}}\longrightarrow $ & $\mathbb{C}P^1$ & & ~~~~~~~~~~(1st)\\
 &  $S^1_{\mathbb{H}}$ & $\longrightarrow$ & $\mathbb{H}P^{1}$ &  & & ~~~~~~~~~~(2nd) \\
 $S^1_{\mathbb{O}}$ & $\longrightarrow$ &   $\mathbb{O}P^{1}$ &&  & & ~~~~~~~~~~(3rd) \\  
\end{tabular}\label{compactdivisionhopf}
\end{center}
It would be worthwhile to mention the groups relevant to the Hopf maps. 
The isometries of the base manifolds, $S^2$, $S^4$ and $S^8$, are respectively given by  
\begin{align}
&SO(3)\simeq SU(2)/Z_2\simeq USp(2)/Z_2,\nonumber\\
&SO(5)\simeq USp(4)/Z_2,\nonumber\\
&SO(9),
\label{isometrycompact}
\end{align}
where $USp(2n)$ denotes the compact symplectic group. 
Similarly, the holonomy groups of the base manifolds are 
\begin{align}
&SO(2)\simeq U(1)\simeq S^1,\nonumber\\
&SO(4)\simeq SU(2)\times SU(2)\simeq S^3\times S^3,\nonumber\\
&SO(8)\simeq S^7\times S^7 \times G_2.
\label{holonomycompact}
\end{align}
It should be noticed that the holonomy groups and the sphere-bundles are closely related:  The sphere-bundles,  $S^1$, $S^3$ and $S^7$,  appear on RHS in the expression of the holonomy groups 
(\ref{holonomycompact}). 
 The 1st and 2nd Hopf maps are special in the sense that their sphere-bundles have counterparts of  the (special) unitary group manifolds, $i.e.$ $U(1)\simeq S^1$, $SU(2)\simeq S^3$. (There does not exist a corresponding group manifold with $S^7$.)
 The automorphism groups of the division algebras are respectively given by 
\begin{equation}
Aut(\mathbb{C})=Z_2,~~~Aut(\mathbb{H})=SO(3),~~~Aut(\mathbb{O})=G_2.
\end{equation}

Few years after the discovery of quaternions by William R. Hamilton \cite{Hamilton1844}, James Cockle introduced the notion of the split-algebras \cite{Cockle1848,Cockle1849}.  
Similar to the original division algebras, there exist three species of split-algebras; split-complex numbers $\mathbb{C}'$, split-quaternions $\mathbb{H}'$, and split-octonions $\mathbb{O}'$. The split-algebras have the properties similar to the original division algebras except for their split signatures \footnote{There also exists another type of quaternions called hyperbolic quaternions introduced by Alexander Macfarlane \cite{Macfarlane1891} corresponding to Lorentzian signature.
However, the hyperbolic quaternions do not respect associativity and it is not probable to construct another Hopf map based on hyperbolic quaternions.}. 
Based on the split-algebras, we introduce a notion of non-compact Hopf maps 
\begin{center}
\begin{tabular}{ccccccc}
\\ 
 & & $H^1_{\mathbb{C'}}$ &   $\overset{H^1_{\mathbb{R}}}\longrightarrow $ & $\mathbb{C'}P^1$ & & ~~~~~~~~~~(1st)\\
 &  $H^1_{\mathbb{H'}}$ & $\longrightarrow$ & $\mathbb{H'}P^{1}$ &  & & ~~~~~~~~~~(2nd) \\
 $H^1_{\mathbb{O'}}$ & $\longrightarrow$ &   $\mathbb{O'}P^{1}$ &&  & & ~~~~~~~~~~(3rd) \\  
\end{tabular}
\end{center}
where $H^1$ represents ``one-dimensional'' hyperboloid (hyperbola) in the space of the 
corresponding split-algebras.
Defining higher dimensional hyperboloids (ultra-hyperboloids)  $H^{p,q}$ \footnote{In particular, 
$S^n=H^{0,n}$, 
$dS^n=H^{1,n-1}$, and $AdS^n=H^{n-1,1}$.}: 
\begin{equation}
  \sum_{i=1}^{p} x^i x^i-\sum_{j=1}^{q+1} y^j y^j=-1,
\label{defofultrahyper}
\end{equation}
the non-compact Hopf maps can be expressed as 
\begin{center}
\begin{tabular}{ccccccc}
\\ 
 & & $H^{2,1}$ &   $\overset{H^{1,0}}\longrightarrow $ & $H^{1,1}$ & & ~~~~~~~~~~(1st)\\
 &  $H^{4,3}$ & $\longrightarrow$ & $H^{2,2}$ &  & & ~~~~~~~~~~(2nd) \\
 $H^{8,7}$ & $\longrightarrow$ &   $H^{4,4}$ &&  & & ~~~~~~~~~~(3rd) \\  
\end{tabular}
\end{center}
The relevant groups of the non-compact Hopf maps are summarized as follows. 
The isometries of the base manifolds, $H^{1,1}$, $H^{2,2}$ and $H^{4,4}$, are 
\begin{align}
&SO(2,1)\simeq SU(1,1)/Z_2\simeq Sp(2,R)/Z_2,\nonumber\\
&SO(3,2)\simeq Sp(4,R)/Z_2,\nonumber\\
&SO(5,4),
\label{isometrynoncompact}
\end{align}
where $Sp(2n,R)$ represents the real symplectic group.
The holonomy groups of the base manifolds are respectively 
\begin{align}
&SO(1,1)\simeq \mathcal{U}(1)\simeq H^{1,0},\nonumber\\
&SO(2,2)\simeq SU(1,1)\times SU(1,1)\simeq H^{2,1}\times H^{2,1},\nonumber\\
&SO(4,4)\simeq H^{4,3}\times H^{4,3} \times G_{2(2)}.
\label{holonomynoncompact}
\end{align}
(The group $\mathcal{U}(1)$ will be defined in Subsec.{\ref{subsecsplitcomp}}.)
On RHS in the expression of the non-compact holonomy groups (\ref{holonomynoncompact}),  one may notice that there appear the  hyperboloid-bundles, $H^{1,0}$, $H^{2,1}$ and $H^{4,3}$.  The 1st and 2nd non-compact Hopf maps are special in the sense that their hyperboloid-bundles have the counterparts of the non-compact (special) unitary group manifolds, $i.e.$ $\mathcal{U}(1)\simeq H^{1,0}$, $SU(1,1)\simeq H^{2,1}$. (There does not exist a corresponding group manifold with $H^{4,3}$.)
The automorphism groups of the split-algebras are given by 
\begin{equation}
Aut(\mathbb{C'})=Z_2,~~~
Aut(\mathbb{H'})=SO(2,1),~~~Aut(\mathbb{O'})=G_{2(2)}.
\end{equation}
Apparently, there are close analogies between the relevant groups of  the original  and the non-compact Hopf maps. 

In this paper, we explore an explicit realization of the non-compact Hopf maps. 
We introduce two ways of the realization, one of which (referred as ``Realization I'' in the context) is performed by using the split-imaginary unit $j$: $j^2=1$, 
and the other (referred as ``Realization II'') by the ordinary imaginary unit $i$: $i^2=-1$. 
The  generators of non-compact groups are represented by  
 finite-dimensional ``hermitian'' matrices, 
 with use of the split-imaginary unit. Taking advantage of this property, in Realization I, we demonstrate the realization of the non-compact Hopf maps quite analogously to that of the original Hopf maps \footnote{The realization of the original Hopf maps can be referred to Refs.\cite{Hopf1931,Hopf1935,ZhangScience2001,Bernevig2003}}.  
In Realization II, we adopt the ordinary imaginary unit, and realize the non-compact Hopf maps. 
 There, finite-dimensional non-hermitian matrices are
 utilized to express the non-compact generators.  
Topological structures of hyperboloid bundles are also explored.
Physically, the canonical connection of each of three non-compact Hopf maps is interpreted as $SO(1,1)$, $SU(1,1)$ and $SO(4,4)$ monopole gauge field. 

The organizations are as follows. 
In Sec.\ref{sec0thnoncompac}, as a preliminary, we introduce 0th non-compact Hopf map.
In Sec.\ref{sec1stnoncompac}, Sec.\ref{sec2ndnoncompac} and Sec.\ref{sec3rdnoncompac}, we develop the non-compact version of the 1st, 2nd and 3rd Hopf maps, respectively.  
The $SO(2,1)$, $SO(3,2)$ Dirac, and $SO(5,4)$ Majorana spinors play crucial roles
  in constructing the non-compact Hopf maps.   
Sec.\ref{secsummary} is devoted to summary and discussions.
In Appendix \ref{secsusynon}, the non-compact supersymmetric (1st) Hopf map is provided.

\section{0th Non-compact Hopf Map}\label{sec0thnoncompac}

As a warm-up, we first introduce the non-compact version of the 0th Hopf map. 
The original 0th Hopf map is given by 
\begin{equation}
S^1 \overset{Z_2}\longrightarrow S^1,
\label{0thcopacthopf}
\end{equation}
where 1D sphere $S^1$, $i.e.$, circle is defined as usual: 
\begin{equation}
{x_1}^2+{x_2}^2=1.
\end{equation}
The map (\ref{0thcopacthopf}) is realized by identifying ``opposite points'' on the circle: 
\begin{equation}
(x_1,x_2)\sim -(x_1,x_2).
\label{identifyz2}
\end{equation}
We introduce the non-compact version of the 0th Hopf map  
\begin{equation}
H^{1,0}\overset{Z_2}\longrightarrow H^{1,0},
\label{0noncompacthopf}
\end{equation}
where $H^{1,0}$ is  one-dimensional hyperboloid, $i.e.$, hyperbola defined by 
\begin{equation}
{x_1}^2-{x_2}^2=-1,
\label{hyperboladef}
\end{equation}
and the hyperbola consists of two branches; left and right in $(x_2,x_1)$-space. 
The non-compact 0th Hopf map is realized by identifying  the ``opposite points'' (with respect to the origin) between the right and left branches of the hyperbola as indicated by Eq.(\ref{identifyz2}), and 
the $H^{1,0}$ on RHS of (\ref{0noncompacthopf}) represents one-branch of hyperbola. 
 The 0th non-compact Hopf map is explicitly given by 
\begin{equation}  
  (x_1,x_2) \rightarrow (y_1,y_2)=(2x_1x_2,x_1^2+x_2^2). 
\end{equation}  
One immediately finds that $y_1$ and $y_2$ satisfy the hyperbola condition,  $y_1^2-y_2^2=-(x_1^2-x_2^2)^2=-1$.  

\section{1st Non-compact Hopf Map}\label{sec1stnoncompac}

The 1st non-compact Hopf map is introduced as  
\begin{equation}
H^{2,1}\overset{H^{1,0}}\longrightarrow H^{1,1},
\label{non-compactHopf1-1}
\end{equation}
or 
\begin{equation}
AdS^3 \overset{\mathcal{U}(1)}\longrightarrow AdS^2.
\label{noncompact1strewritten}
\end{equation}
Here, $AdS^n$ denotes the $n$-dimensional anti de Sitter space, and the base manifold  $AdS^2\simeq H^{1,1}$ is a two-dimensional one-leaf hyperboloid.  

\subsection{Split-Complex Numbers}\label{subsecsplitcomp}

The split-imaginary unit $j$ is introduced so as to satisfy  
\begin{equation}
j^2=1, ~~~~~~j^*=-j,
\label{propertyofj}
\end{equation}
where $*$ denotes complex conjugation. 
With two real numbers $x$ and $y$, the spilt-complex number is defined as  
\begin{equation}
z=x+jy,
\end{equation}
and its complex conjugation is  
\begin{equation}
z^*=x-jy.
\end{equation}
Then, we find 
\begin{equation}
z^*z=zz^*=x^2-y^2. 
\end{equation}
The ``normalized''split-complex number satisfying $z^*z=zz^*=-1$  represents $H^{1,0}$  \footnote{With existence of negative signature, positive norm (or more generally magnitude) is not properly defined. However, for brevity, in the present paper, we adopt the terminology ``normalized'' even for spinor whose self-inner product is -1.} .  
The hyperbola $H_{\mathbb{C}'}^{1}$ in the 2D split-complex space is defined as 
\begin{equation}
z^*z-{z'}^*z'=x^2-y^2-{x'}^2+{y'}^2=-1,
\end{equation}
and thus $H_{\mathbb{C}'}^{1}$ represents $H^{2,1}$. 

Utilizing the split-imaginary unit $j$ instead of $i$, we define a  $\it{non}$-$\it{compact}$ $\mathcal{U}(1)$  group whose  element is given by    
\begin{equation}
e^{j\vartheta}=\cosh \vartheta +j\sinh \vartheta,
\end{equation}
which has one-to-one correspondence to the $SO(1,1)$ group element 
\begin{equation}
g=
\begin{pmatrix}
\cosh \vartheta & \sinh \vartheta \\
\sinh \vartheta & \cosh \vartheta
\end{pmatrix}.
\end{equation}
Therefore, the $\mathcal{U}(1)$ group is isomorphic to the $SO(1,1)$ group
\begin{equation}
\mathcal{U}(1)\simeq SO(1,1). 
\end{equation}
With real numbers $x$ and $y$, the $SO(1,1)$ group element is generally represented as  
$g=\begin{pmatrix}
x & y \\
y & x
\end{pmatrix}$ satisfying the constraint $\text{det}(g)=x^2-y^2=1$. This 
 constraint coincides with the definition of  $H^{1,0}$, and then 
\begin{equation}
\mathcal{U}(1) \simeq H^{1,0}.
\end{equation}

\subsection{Realization I}\label{splitimaginaryhopf1}

First, we introduce the spilt-Pauli matrices   
\begin{equation}
\sigma^1=
\begin{pmatrix}
0 & 1 \\
1 & 0
\end{pmatrix},~~~
\sigma^2=
\begin{pmatrix}
0 & -j \\
j & 0
\end{pmatrix},~~~
\sigma^3=
\begin{pmatrix}
1 & 0 \\
0 & -1
\end{pmatrix}.
\label{splitimaginaryPauli}
\end{equation}
The split-imaginary unit $j$ appears in $\sigma^2$, and the split Pauli matrices satisfy the anticommutation relations 
\begin{equation} 
 \{\sigma^i,\sigma^j\}=2\eta^{ij},
\end{equation}
where $\eta^{ij}=\eta_{ij}=\text{diag}(+1,-1,+1)$. The split-Pauli matrices can also be regarded as the gamma matrices in (2+1)D.   
The split-Pauli matrices are hermitian and satisfy the $SU(1,1)$ algebra 
\begin{equation}
[\sigma^i,\sigma^j]=2j\epsilon^{ij}_{~~k}\sigma^k, 
\end{equation}
with $\epsilon^{123}=1$. 
 It is noted that, in general, generators of non-compact groups cannot be expressed by finite-dimensional hermitian matrices, but here, because of the special property of the split-imaginary unit $j$ (\ref{propertyofj}), the $SU(1,1)$ generators can be represented by hermitian and finite-dimensional matrices. 
Since $\sigma^2$ satisfies 
\begin{equation}
-(\sigma^i)^*=\sigma^2 \sigma^i (\sigma^2)^{-1}, 
\end{equation}
$\sigma^2$ is the charge conjugation matrix of $SU(1,1)$, and has the following properties  
\begin{equation}
(\sigma^2)^{-1}=-\sigma^2= (\sigma^2)^t=(\sigma^2)^*. 
\end{equation}
The consistency condition is satisfied as  
\begin{equation}
(\sigma^2)^*\sigma^2=\sigma^2 (\sigma^2)^*=1_2.
\end{equation}
(Here, $1_2$ denotes $2\times 2$ unit matrix.) 

With use of the split Pauli matrices (\ref{splitimaginaryPauli}), the 1st non-compact Hopf map is expressed as  
\begin{equation}
\phi\rightarrow x^i=\phi^{\dagger}\sigma^i \phi.
\label{explicitnon1stsplit}
\end{equation}
Here, $\phi$ denotes a $SU(1,1)$ Dirac spinor (the non-compact 1st Hopf spinor)  
\begin{equation}
\phi=
\begin{pmatrix}
u\\
v
\end{pmatrix}
\end{equation}
 subject to the ``normalization condition''  
\begin{equation}
\phi^{\dagger}\phi=u^*u+v^*v={u_R}^2-{u_I}^2+{v_R}^2-{v_I}^2=1,
\label{normalizationphiI}
\end{equation}
where $u=u_R+ju_I$, $v=v_R+jv_I$. (The lower indices $R$ and $I$ represent real and imaginary parts, respectively.)  
Thus, $\phi$ denotes  coordinates on  $H^{2,1}$.  
The components of the $SU(1,1)$ vector $x^i$  (\ref{explicitnon1stsplit}) are given by  
\begin{equation}
x^1=u^*v+v^*u,~~~x^2=-ju^*v+jv^*u,~~~x^3=u^*u-v^*v,  
\end{equation}
and they satisfy 
\begin{equation}
\sum_{i,j=1,2,3}\eta_{ij}x^ix^j=(x^1)^2-(x^2)^2+(x^3)^2=(\phi^{\dagger}\phi)^2=1.
\end{equation}
Then, $x^i$ denote coordinates on $H^{1,1}$. 

On the upper patch of  $H^{1,1}$ ($x^3\ge0$), we invert the map (\ref{explicitnon1stsplit}) to represent $\phi$ as  
\begin{equation}
\phi=\frac{1}{\sqrt{2(1+x^3)}}
\begin{pmatrix}
1+x^3\\
x^1-jx^2
\end{pmatrix}
.\label{formofsplithopfspi1}
\end{equation}
The canonical connection of $H^{1,0}$-fibre is induced as 
\begin{equation}
A=dx^i A_i=-j\phi^{\dagger}d\phi,
\end{equation}
with   
\begin{equation}
A_i=\frac{1}{2(1+x^3)}\epsilon_{ij3}x^j.
\label{u1connection1}
\end{equation}
The curvature is derived as     
\begin{equation}
F=dA=\frac{1}{2}dx^i \wedge dx^j F_{ij}
\end{equation}
where 
\begin{equation}
F_{ij}=\partial_i A_j-\partial_j A_i=-\frac{1}{2}\epsilon_{ijk}x^k.
\label{splitcompfieldstregths}
\end{equation}

Similarly, on the lower patch ($x^3\le 0$), the Hopf spinor satisfying (\ref{explicitnon1stsplit}) is given by 
\begin{equation}
\phi'=\frac{1}{\sqrt{2(1-x^3)}}
\begin{pmatrix}
x^1+jx^2\\
1-x^3
\end{pmatrix}.
\end{equation}
The corresponding canonical connection is 
\begin{equation}
A'_i=-\frac{1}{2(1-x^3)}\epsilon_{ij3}x^j,
\label{u1connection2}
\end{equation}
and the curvature $F'_{ij}=\partial_i A'_j-\partial_j A'_i$ is the same (\ref{splitcompfieldstregths}). 
The $\mathcal{U}(1)$ transition function which connects the fibres on two patches is given by  
\begin{equation}
g=\frac{x^1+jx^2}{\sqrt{1-(x^3)^2}}. 
\label{transj1st}
\end{equation}
It satisfies the condition $g^{*}g=gg^{*}=1$. 
The function $g$ is the map from a point on $H^{1,0}$ to a group element of  $\mathcal{U}(1)$, since the coordinates $\hat{x}^i=\frac{1}{\sqrt{1-(x^3)^2}}x^i$ ($i=1,2$) in $g$ satisfy $\hat{x}^1\hat{x}^1-\hat{x}^2\hat{x}^2=-1$ and represent $H^{1,0}$. 
The manifold $H^{1,0}$ is the ``equator'' of the base manifold $H^{1,1}$,  and the transition function (\ref{transj1st}) specifies  the ``gluing'' of the $\mathcal{U}(1)$ fibres on the upper and lower  patches on the ``equator'' of the base manifold, as usual. 
The differential of $g$ yields
\begin{equation}
-jg^* dg=-jdg g^*=-\frac{1}{1-(x^3)^2}\epsilon_{ij3}x^jdx^i.
\end{equation}
Then, the canonical connections (\ref{u1connection1}) and (\ref{u1connection2}) are expressed as 
\begin{align}
&A_i dx^i=j\frac{1}{2}(1-x^3)g^*dg,\nonumber\\
&A'_i dx^i=-j\frac{1}{2}(1+x^3)g^*dg, 
\end{align}
and  we confirm the relation 
\begin{equation}
A'_i dx^i=A_i dx^i-jg^*dg,
\end{equation}
and 
\begin{equation}
F_{ij}'=F_{ij}. 
\end{equation}

Here we add some comments. Introducing a radial coordinate $r$ in the hyperbolic space as 
\begin{equation}
r^2={\eta_{ij}x^i x^j}=(x^1)^2-(x^2)^2+(x^3)^2,
\end{equation}
the curvature (\ref{splitcompfieldstregths}) becomes 
\begin{equation}
F_{ij}=-\frac{1}{2r^3}\epsilon_{ijk}x^k.
\end{equation}
This expression is superficially equivalent to that of the Dirac monopole, but there is a crucial difference. The location of monopole corresponds to the singularity point $r=0$. For the ordinary Dirac monopole, such condition can  be satisfied only at the origin $x^i=0$, since the space is Euclidean. However, for the non-compact monopole, due to the hyperbolic signature,  such condition can be  satisfied at any point on  surface of light-cone:  $(x^1)^2-(x^2)^2+(x^3)^2=0$. Therefore, the ``localization'' of the non-compact monopole may be understood as the surface of the light-cone.

\subsection{Realization II}\label{ordinaryimaginary1stHopf}

In the case of the ordinary imaginary unit, 
the non-compact fibre $\mathcal{U}(1)\simeq H^{1,0}$  is replaced by the compact fibre $U(1)\simeq H^{0,1}$, and then the 1st non-compact Hopf map takes the different form from (\ref{non-compactHopf1-1}):   
\begin{equation}
H^{2,1}\overset{H^{0,1}}\longrightarrow H^{2,0},
\label{differentnoncompacthopfmap}
\end{equation}
or 
\begin{equation}
AdS^3 \overset{{U}(1)}\longrightarrow \text{Euclid.}AdS^2.
\label{1stmapsrewritten}
\end{equation}
$\text{Euclid.}AdS^2$ is a two-leaf hyperboloid, while $AdS^2$  in (\ref{noncompact1strewritten}) is a connected (one-leaf) hyperboloid.  

The $SU(1,1)(\simeq SO(2,1))$ generators $\tau^i$ are realized as  
\begin{equation}
\tau^1=i\sigma^1,~~~\tau^2=i\sigma^2,~~~\tau^3=\sigma^3,
\label{defoftaus}
\end{equation}
where  $\sigma^i$ denote the ordinary Pauli matrices 
\begin{equation}
\sigma^1=
\begin{pmatrix}
0 & 1 \\
1 & 0
\end{pmatrix},~~~
\sigma^2=
\begin{pmatrix}
0 & -i \\
i & 0
\end{pmatrix},~~~
\sigma^3=
\begin{pmatrix}
1 & 0 \\
0 & -1
\end{pmatrix}.
\label{imaginaryPauli}
\end{equation}
The matrices $\tau^i$ are non-hermitian and satisfy 
\begin{equation}
(\tau^i)^{\dagger}=-\tau_i.
\end{equation}
In detail, $\tau^1$ and $\tau^2$ are anti-hermitian, while $\tau^3$ is hermitian. 
The matrices $\tau^i$ satisfy the anticommutation relations 
\begin{equation}
\{\tau^i,\tau^j\}=-2\eta^{ij},
\end{equation}
with $\eta^{ij}=\eta_{ij}=diag(+,+,-)$, and their commutation relations are 
\begin{equation}
[\tau^i,\tau^j]=2i\epsilon^{ijk}\tau_k.
\end{equation}
The charge conjugation matrix $\sigma^1$ satisfies  
\begin{equation}
-(\tau^i)^*=\sigma^1 \tau^i (\sigma^1)^{-1},  
\end{equation}
and has the properties 
\begin{equation}
(\sigma^1)^{-1}=(\sigma^1)^t=(\sigma^1)^{*}=(\sigma^1)^{\dagger}.
\end{equation}
By a unitary transformation, $\sigma^1$ can be diagonalized to give $\sigma^3$. Multiplying 
$\sigma^3$ to $\tau^i$,  hermitian matrices are constructed as 
\begin{equation}
\sigma^3\tau^i=(-\sigma^2,\sigma^1,1).
\end{equation}

A ``normalized'' $SO(2,1)$ Dirac spinor $\phi=(u,v)^t$ (1st non-compact Hopf spinor) is introduced as 
\begin{equation}
\phi^{\dagger}\sigma^3\phi=u^*u-v^*v=u_R^2+u_I^2-v_R^2-v_I^2=1,
\label{normalphiII}
\end{equation}
and  $\phi$ represents the coordinates on $H^{2,1}$. 
We now realize the 1st non-compact Hopf map (\ref{differentnoncompacthopfmap}) explicitly  
\begin{equation}
\phi 
\rightarrow 
x^i= \phi^{\dagger}  \sigma^3 \tau^i\phi,
\label{1stHopfmapexplicit}
\end{equation}
or
\begin{equation}
x^1=iu^*v-iv^*u,~~~x^2=u^*v+v^*u,~~~x^3=u^*u+v^*v, 
\end{equation}
which denote coordinates on $H^{2,0}$ as found  
\begin{equation}
\sum_{i,j=1,2,3}\eta_{ij}x^i x^j=x^2+y^2-z^2=-(\phi^{\dagger}\sigma^3\phi)^2=-1.
\end{equation}
On the upper leaf of $H^{2,0}$ $(x^3\ge 1)$, the 1st non-compact Hopf spinor can be expressed as  
\begin{equation}
\phi=\frac{1}{\sqrt{2(1+x^3)}}
\begin{pmatrix}
1+x^3\\
x^2-ix^1
\end{pmatrix}.
\end{equation}
The associated canonical connection is evaluated as 
\begin{equation}
A=-i\phi^{\dagger} \sigma^3 d\phi=dx^i A_i,
\end{equation}
with  
\begin{equation}
A_i=-\frac{1}{2(1+x^3)}\epsilon_{ij3}{x^j}.
\label{u1conne1}
\end{equation}
Similarly, the curvature is derived  
\begin{equation}
F_{ij}=\partial_i A_j-\partial_j A_i=\frac{1}{2}\epsilon_{ijk}{x^k}.
\end{equation}
A non-singular canonical connection on the lower leaf $(x^3\le -1)$ can be given by 
\begin{equation}
A'_i=\frac{1}{2(1-x^3)}\epsilon_{ij3}x^j,
\label{u1conne2}
\end{equation}
and the corresponding curvature is  
\begin{equation}
F'_{ij}=F_{ij}. 
\end{equation}
Unlike the canonical connection in Realization I, the present canonical connection on two-leaf hyperboloid may be topologically trivial, since each leaf is topologically equivalent to a  2D-plane.

\section{2nd Non-compact Hopf Map}\label{sec2ndnoncompac}

In this section, based on the split-quaternions, we explore a realization of the 2nd non-compact Hopf map: 
\begin{equation}
H^{4,3} \overset{H^{2,1}}\longrightarrow H^{2,2}. 
\label{noncompact2ndhopfh}
\end{equation}

\subsection{Split-Quaternions}

The split-quaternions, $1,q_1,q_2,q_3$,  are introduced so as to satisfy the following algebras:    
\begin{align}
&{q_1}^2=-{q_2}^2={q_3}^2=q_1q_2q_3=1,\nonumber\\
&q_iq_j=-q_j q_i~~(i\neq j),
\label{relationquaternion}
\end{align}
with $i,j=1,2,3$. 
As in the case of  quaternions, the commutativity does not hold in the algebra of the split-quaternions.
Other relations such as 
\begin{equation}
q_1q_2=q_3,~~q_2 q_3=q_1,~~q_3 q_1=-q_2, 
\label{otherrelaquaternion}
\end{equation}
can be derived by (\ref{relationquaternion}). 
The conjugation of the split-quaternions is given by 
\begin{equation}
1^*=1,~~~~{q_i}^*=-q_i.
\label{complexpropertyofquaternion}
\end{equation}
A split-quaternion is generally represented as 
\begin{equation}
h=r_0 1+r_1 q_1 +r_2 q_2 +r_3 q_3, 
\end{equation}
and its conjugation becomes  
\begin{equation}
h^*=r_0 1-r_1 q_1 -r_2 q_2 -r_3 q_3. 
\end{equation}
Then, we  find 
\begin{equation}
h^*h=hh^*= {r_0}^2-{r_1}^2+{r_2}^2-{r_3}^2.
\end{equation}
Thus, the self-inner product of split-quaternions possesses the split-signatures, two of which are positive and the other two are negative.  
The ``normalized'' split-quaternion is defined so as to satisfy $h^*h=hh^*=-1$, which represents $H^{2,1}$.
Similarly, the hyperbola $H_{\mathbb{H}'}^{1}$ in the 2D split-quaternion space is defined as
\begin{equation}
h^* h-h'^* h   = {r_0}^2-{r_1}^2+{r_2}^2-{r_3}^2- {r'_0}^2+{r'_1}^2-{r'_2}^2+{r'_3}^2=-1,
\end{equation} 
and thus, $H_{\mathbb{H}'}^{1}$ is equal to  $H^{4,3}$.

It is also important to notice that the split-quaternions are realized with use of the split-Pauli matrices: 
\begin{equation}
q_i=j\sigma^i,
\end{equation}
where $j$ in front of $\sigma^i$ are needed to satisfy Eq.(\ref{complexpropertyofquaternion}) 
with the definition  $(q_i)^*= (j\sigma^i)^{\dagger}$.

\subsection{Realization I}\label{subsucrealizationIfornoncompact2nd}

The $SO(3,2)$ anticommutation relations are given by 
\begin{equation}
\{\gamma^a,\gamma^b\}=-2\eta^{ab}, 
\end{equation}
with $\eta^{ab}=\eta_{ab}=\text{diag}(+1,-1,+1,-1,-1)$.  
The $SO(3,2)$ gamma matrices  $\gamma^a=(\gamma^{i},\gamma^4,\gamma^5)$ ($i=1,2,3$) can be represented as 
\begin{equation}
\gamma^i=
\begin{pmatrix}
0 & j\sigma^i \\
-j\sigma^i & 0
\end{pmatrix},~~
\gamma^4=
\begin{pmatrix}
0 & 1_2 \\
1_2 & 0
\end{pmatrix},~~
\gamma^5=
\begin{pmatrix}
1_2 & 0 \\
0 & -1_2
\end{pmatrix}.
\label{so32gammaspilitim}
\end{equation}
The ``split-quaternions'' appear in the off-diagonal elements of $\gamma^a$. Just as in the case of $SU(1,1)$, with the use of $j$,  the $SO(3,2)$ gamma matrices are represented by hermitian matrices: $\gamma^1$, $\gamma^3$ are pure imaginary antisymmetric matrices, and $\gamma^2$, $\gamma^4$ and $\gamma^5$ are real symmetric matrices. Then, the $SO(3,2)$ generators constructed by the formula  
$\sigma^{ab}=-j\frac{1}{4}[\gamma^a,\gamma^b]$ are also represented by hermitian and finite dimensional matrices. 
 The charge conjugation matrix of $SO(3,2)$ is constructed by multiplying the purely imaginary gamma matrices,   
\begin{equation}
b=j\gamma^1\gamma^3=
-
\begin{pmatrix}
\sigma^2 & 0 \\
0 & \sigma^2
\end{pmatrix}
=
j\begin{pmatrix}
\epsilon & 0 \\
0 & \epsilon 
\end{pmatrix}, 
\label{chargeconjSO(3,2)}
\end{equation}
where 
\begin{equation}
\epsilon=\epsilon_{\alpha\beta}=\begin{pmatrix}
0 & 1 \\
-1 & 0
\end{pmatrix}. 
\end{equation}
Indeed, the matrix $b$ (\ref{chargeconjSO(3,2)}) satisfies 
\begin{equation}
b\gamma_a b^{-1}=\gamma_a^*,~~b\sigma_{ab}b^{-1}=-\sigma_{ab}^*,
\end{equation}
and 
\begin{equation}
b^{-1}=-b=b^{t}=b^*.
\end{equation}
The consistency condition is satisfied,  
\begin{equation}
b^*b=bb^*(=-b^2)=1.
\end{equation}
The charge conjugation  of $\psi=(\psi_1,\psi_2,\psi_3,\psi_4)^t$ is constructed as 
\begin{equation}
\psi_c=-b^{-1}\psi^*=b\psi^*, 
\label{chargeconjspinorSO(3,2)}
\end{equation}
and its self-inner product becomes  
\begin{equation}
\psi_c^{\dagger}\psi_c=-\psi^{\dagger}\psi.
\end{equation}
With the above preparation, we now realize the 2nd non-compact Hopf map  as  
\begin{equation}
\psi\rightarrow x^a=\psi^{\dagger}\gamma^a \psi,   
\label{noncompact2nghopdfmapII}
\end{equation}
where  $\psi$ denotes a ``normalized'' $SO(3,2)$ spinor subject to the condition 
\begin{equation}
\psi^{\dagger}\psi=\sum_{\mu=1}^4\psi_{\mu}^*\psi_{\mu}=
\sum_{\mu=1}^4 (\psi_R^{\mu})^2-\sum_{\mu=1}^4 (\psi_I^{\mu})^2=1,
\label{normalization2ndHopfspinor}
\end{equation}
with $\psi^{\mu}=\psi_R^{\mu}+j\psi^{\mu}_I$. The condition (\ref{normalization2ndHopfspinor}) is the definition of $H^{4,3}$, and then $\psi$ represents  coordinates on $H^{4,3}$. Meanwhile, the explicit formulas for $x^a$ (\ref{noncompact2nghopdfmapII}) are given by 
\begin{align}
&x^1=-j(\psi_4^*\psi_1+\psi_3^*\psi_2-\psi_2^*\psi_3-\psi_1^*\psi_4),\nonumber\\
&x^2=-\psi_4^*\psi_1+\psi_3^*\psi_2+\psi_2^*\psi_3-\psi_1^*\psi_4,\nonumber\\
&x^3=-j(\psi_3^*\psi_1-\psi_4^*\psi_2-\psi_1^*\psi_3+\psi_2^*\psi_4),\nonumber\\
&x^4=\psi_3^*\psi_1+\psi_4^*\psi_2+\psi_2^*\psi_3+\psi_2^*\psi_4,\nonumber\\
&x^5=\psi_1^*\psi_1+\psi_2^*\psi_2-\psi_3^*\psi_3-\psi_4^*\psi_4,
\end{align}
and satisfy the relation
\begin{equation}
\sum_{a,b=1,2,3,4,5}\eta_{ab}x^a x^b=(x^1)^2-(x^2)^2+(x^3)^2-(x^4)^2-(x^5)^2=-(\psi^{\dagger}\psi)^2=-1.
\end{equation}
Thus, $x^a$ represent coordinates on $H^{2,2}$. 

Inverting the 2nd non-compact Hopf map (\ref{noncompact2nghopdfmapII}) on the upper patch of  $H^{2,2}$ $(x^5\ge 0)$,  the non-compact Hopf spinor can be represented as 
\begin{equation}
\psi= \frac{1}{\sqrt{2(1+x^5)}}
\begin{pmatrix}
(1+x^5)\phi \\ 
(x^4+jx^i\sigma_i)\phi
\end{pmatrix},
\label{explicit2ndHopfspinorII}
\end{equation}
where $\phi$ is a ``normalized'' 2-component spinor (\ref{normalizationphiI}) representing the $AdS^3$-fibre. 
A straightforward calculation shows that  $\phi$ is cancelled in the map (\ref{noncompact2nghopdfmapII}). 
Utilizing the explicit form (\ref{explicit2ndHopfspinorII}), the canonical connection of $AdS^3$-fibre is evaluated as 
\begin{equation}
A=-j\psi^{\dagger}d\psi=dx^a \phi^{\dagger}A_a\phi,
\end{equation}
where 
\begin{align}
&A_{m}=-\frac{1}{2(1+x^5)} \eta_{mn i} x^{n}\sigma^i,\nonumber\\
&A_5=0.
\label{su11conne1}
\end{align}
Here, $m,n=1,2,3,4$,  and  $\eta_{mn i}$ is the split-'t Hooft symbol made of the split-metric $\eta_{mn}=\text{diag}(+1,-1,+1,-1)$ as  
\begin{equation}
\eta_{mn i}=\epsilon_{mn i}-\eta_{m i}\eta_{n 4}+\eta_{m 4}\eta_{n i}. 
\end{equation}
The curvature
\begin{equation}
F=dA -j A\wedge A=\frac{1}{2}dx^a\wedge dx^b F_{ab}
\end{equation}
or 
\begin{equation}
F_{ab}=\partial_a A_b-\partial_{b}A_a-j[A_a,A_b]
\end{equation}
is  computed as 
\begin{align}
&F_{mn}=x_{m}A_{n}-x_{n}A_{m}+ \frac{1}{2}\eta_{mn i}\sigma^i,\nonumber\\
&F_{m 5}=-F_{5m}=-(1+x^5)A_{m}.
\end{align}

In the lower patch ($x^5\le 0$), the 2nd non-compact Hopf spinor can be expressed as 
\begin{equation}
\psi'=\frac{1}{\sqrt{2(1-x^5)}}
\begin{pmatrix}
(x^4-jx^i\sigma_i)\phi\\
(1-x^5)\phi 
\end{pmatrix},
\label{2ndlownonspiII}
\end{equation}
and the canonical connection is derived as 
\begin{equation}
A'=-j\psi'^{\dagger}d\psi=dx^a \phi^{\dagger}A'_a \phi
\end{equation}
where 
\begin{align}
&A'_{m}=-\frac{1}{2(1-x^5)} \bar{\eta}_{mn i} x^{n}\sigma^i,\nonumber\\
&A'_5=0,
\label{su11conne2}
\end{align}
with $\bar{\eta}_{mn i}=\epsilon_{mn i}+\eta_{m i}\eta_{n 4}-\eta_{m 4}\eta_{n i}. $
The corresponding curvature becomes 
\begin{align}
&F'_{mn}=x_{m}A'_{n}-x_{n}A'_{m}+ \frac{1}{2}\bar{\eta}_{mn i}\sigma^i,\nonumber\\
&F'_{m 5}=-F_{5m}=(1-x^5)A'_{m}.
\end{align}

The transition function which connects (\ref{explicit2ndHopfspinorII}) and (\ref{2ndlownonspiII}) is given by 
\begin{equation}
g=\frac{1}{\sqrt{1-(x^5)^2}}(x^4-jx^i\sigma_i),
\end{equation}
and it satisfies $g^{\dagger}g=1$. 
The function $g$ is  the map from a point on  $H^{2,1}$ to an   element of the structure group  $SU(1,1)$, since the coordinates  $\hat{x}^{m}=\frac{1}{\sqrt{1-(x^5)^2}}x^m$  in $g$  satisfy $\eta_{mn}{\hat{x}}^m {\hat{x}}^n=-1$ and represent   $H^{2,1}$. The transition function specifies  the ``gluing'' of the  $SU(1,1)$ fibres on upper and lower patches  on the ``equator'' $H^{2,1}$ of the base manifold $H^{2,2}$.    
The differential of $g$ yields 
\begin{align}
&-jg^{\dagger}dg=-\frac{1}{1-(x^5)^2}\bar{\eta}_{mni}x^n dx^m \sigma^i,\nonumber\\
&-jdg g^{\dagger}=\frac{1}{1-(x^5)^2}{\eta}_{mni}x^n dx^m \sigma^i, 
\end{align}
and the canonical connections (\ref{su11conne1}) and (\ref{su11conne2}) can be expressed as 
\begin{align}
&A_a dx^a=j\frac{1}{2}(1-x^5)dg g^{\dagger},\nonumber\\
&A'_a dx^a=-j\frac{1}{2}(1+x^5) g^{\dagger}dg. 
\end{align}
Then, we find the canonical connections are related as 
\begin{equation}
A'_a dx^a=g^{\dagger}(A_a dx^a) g-jg^{\dagger}dg, 
\end{equation}
and  the curvatures are 
\begin{equation}
F'_{ab}=g^{\dagger}F_{ab}g.
\end{equation}
Since the structure group is $SU(1,1)$ and the non-trivial transition function is defined, 
the present canonical connection may naturally be interpreted as the $SU(1,1)$ monopole gauge field. 

\subsection{Realization II}

With use of the ordinary imaginary unit, the $(3+2)$D  gamma matrices $\gamma^a$  are realized as  
\begin{equation}
\gamma^i=\tau^i\otimes \sigma^2,~~\gamma^4=1\otimes \sigma^1,~~\gamma^5=\gamma^1\gamma^2\gamma^3\gamma^4 =1\otimes \sigma^3,  
\end{equation}
or 
\begin{align}
&\gamma^i=
\begin{pmatrix}
0 & -i\tau^i\\
i\tau^i & 0
\end{pmatrix},~~\gamma^4=
\begin{pmatrix}
0 & 1_2\\
1_2 & 0
\end{pmatrix},~~\gamma^5=
\begin{pmatrix}
1_2 & 0\\
0 & -1_2
\end{pmatrix}, 
\end{align}
where $i=1,2,3$, and $\tau^i$ are the $SU(1,1)$ generators (\ref{defoftaus}).  
The $SO(3,2)$ gamma matrices $\gamma^a$ satisfy
\begin{equation}
(\gamma^a)^{\dagger}=-\gamma_a.  
\end{equation}
In detail, $\gamma^1$ and $\gamma^2$ are anti-hermitian  while $\gamma^3$, $\gamma^4$ and $\gamma^5$ are hermitian. 
Their anticommutation relations are given by 
\begin{equation}
\{\gamma^a,\gamma^b\}=-2\eta^{ab},
\end{equation}
with $\eta^{ab}=\eta_{ab}=\text{diag}(+,+,-,-,-)$, and 
their commutators yield the $SO(3,2)$ generators \footnote{ The $SO(3,2)$ algebra of $\sigma_{ab}$ 
 is   
$[\sigma_{ab},\sigma_{cd}]=-i(\eta_{ac}\sigma_{bd}-\eta_{ad}\sigma_{bc}+\eta_{bd}\sigma_{ac}-\eta_{bc}\sigma_{ad}).$} 
\begin{equation}
\sigma^{ab}=-i\frac{1}{4}[\gamma^a,\gamma^b],
\end{equation}
or more explicitly 
\begin{align}
&\sigma^{ij}=\frac{1}{2}\epsilon^{ijk}
\begin{pmatrix}
\tau_k & 0 \\
0 & \tau_k
\end{pmatrix},~~
\sigma^{i4}=\frac{1}{2}
\begin{pmatrix}
-\tau^i & 0 \\
0 & \tau^i
\end{pmatrix},\nonumber\\
&\sigma^{i5}=\frac{1}{2}
\begin{pmatrix}
0 & \tau^i \\
\tau^i & 0 
\end{pmatrix},~~
\sigma^{45}=
\frac{i}{2}
\begin{pmatrix}
0 & 1 \\
-1 & 0
\end{pmatrix}. 
\end{align}
The $SO(3,2)$ generators $\sigma^{ab}$ are non-hermitian  
\begin{equation}
(\sigma^{ab})^{\dagger}=\sigma_{ab}.
\end{equation}
The $SO(3,2)$ charge conjugation matrix $r$ is given by 
\begin{equation}
r=-\gamma^2\gamma^3=
\begin{pmatrix}
\sigma^1 & 0 \\
0 & \sigma^1
\end{pmatrix},
\end{equation}
where $\sigma^1$ is the $SO(2,1)$ charge conjugation matrix.  
Indeed, the matrix $r$ satisfies the relations 
\begin{equation}
r^{\dagger}\gamma^a r=(\gamma^a)^{*},~~r^{\dagger}\sigma^{ab}r=-(\sigma^{ab})^*,  
\end{equation}
and has the properties 
\begin{equation}
r^{\dagger}=r^t=r^{-1}=r. 
\end{equation}
Diagonalizing $r$, we obtain  
\begin{equation}
k=
\begin{pmatrix}
\sigma^3 & 0 \\
0 & \sigma^3
\end{pmatrix},
\end{equation}
which satisfies  
\begin{equation}
k^{\dagger}=k^t=k^{-1}=k.
\end{equation}
Multiplying $k$ to $\gamma^a$,  hermitian  matrices are constructed as   
\begin{equation}
k^a=k\gamma^a, 
\end{equation}
or 
\begin{align}
&k^1=
\begin{pmatrix}
0 & i\sigma^2\\
-i\sigma^2 & 0
\end{pmatrix},~~
k^2=
\begin{pmatrix}
0 & -i\sigma^1 \\
i\sigma^1 & 0
\end{pmatrix},~~
k^3 =
\begin{pmatrix}
0 & -i1_2 \\
i1_2 & 0
\end{pmatrix},\nonumber\\
&k^4=
\begin{pmatrix}
0 & \sigma^3 \\
\sigma^3 & 0
\end{pmatrix},~~~~~~~
k^5=
\begin{pmatrix}
\sigma^3 & 0 \\
0 & -\sigma^3
\end{pmatrix}.
\end{align}
Similarly, $k\sigma^{ab}$ are hermitian matrices. 

 A $SO(3,2)$ Dirac spinor $\psi$ (the 2nd non-compact Hopf spinor) subject to the ``normalization condition''  
\begin{equation}
\psi^{\dagger} k\psi=\psi_1^*\psi_1-\psi_2^*\psi_2+\psi_3^*\psi_3-\psi_4^*\psi_4=1,
\end{equation}
 denotes coordinates on  $H^{4,3}$. 
With such a $SO(3,2)$ spinor, the 2nd non-compact Hopf map is realized as   
\begin{equation}
\psi\rightarrow x^a=\psi^{\dagger}k^a\psi.
\label{2ndnoncompactHopf}
\end{equation}
Since $k^a$ are hermitian matrices, the components of the $SO(3,2)$ vector $x^a$ (\ref{2ndnoncompactHopf}) 
\begin{align}
&x^1=\psi_1^*\psi_4-\psi^*_2\psi_3+\psi_4^*\psi_1-\psi_3^*\psi_2,\nonumber\\
&x^2=-i(\psi_1^*\psi_4+\psi^*_2\psi_3-\psi_4^*\psi_1-\psi_3^*\psi_2),\nonumber\\
&x^3=-i(\psi_1^*\psi_3+\psi^*_2\psi_4-\psi_3^*\psi_1-\psi_4^*\psi_2),\nonumber\\
&x^4=\psi_1^*\psi_3-\psi^*_2\psi_4+\psi_3^*\psi_1-\psi_4^*\psi_2,\nonumber\\
&x^5=\psi_1^*\psi_1-\psi^*_2\psi_2-\psi_3^*\psi_3+\psi_4^*\psi_4, 
\end{align}
are real, and  satisfy the relation
\begin{align}
\sum_{a,b=1,2,3,4,5}\eta_{ab}x^a x^b&=(x^1)^2+(x^2)^2-(x^3)^2-(x^4)^2-(x^5)^2\nonumber\\
&=-(\psi^{\dagger}k\psi)^2=-1.
\end{align}
Thus, $x^a$ can be regarded as coordinates on $H^{2,2}$. 
Inverting the 2nd non-compact  Hopf map (\ref{2ndnoncompactHopf}) on the upper patch of the $H^{2,2}$ ($x^5\ge 0$), the  non-compact 2nd Hopf spinor is expressed as  
\begin{equation}
\psi=\frac{1}{\sqrt{2(1+x^5)}}
\begin{pmatrix}
(1+x^5)\phi\\
(x^4-ix^i\tau_i)\phi
\end{pmatrix},
\label{explicit2ndspinor}
\end{equation}
where $\phi=(u,v)^t$ represents the $H^{2,1}$-fibre or the 1st non-compact Hopf spinor that  satisfies the condition (\ref{normalphiII}).  
The canonical connection is evaluated as 
\begin{equation}
A=-i\psi^{\dagger} k d\psi  = dx^a \phi^{\dagger}\sigma^3 A_a \phi 
\end{equation}
where  
\begin{align}
&A_{m}=-\eta_{mn i } \frac{x^{n}}{2(1+x^5)}\tau^i,\nonumber\\
&A_5=0. 
\label{su11connections1}
\end{align}
Here, the split-'t Hooft symbol is given by   
\begin{equation}
\eta_{mn i }=\epsilon_{mn i 4}+ \eta_{m i}\eta_{n 4}-\eta_{n i}\eta_{m 4}.
\end{equation}
It is straightforward to compute 
the curvature by utilizing the formula 
\begin{equation}
F_{ab}=\partial_a A_b-\partial_b A_a+i[A_a,A_b],
\end{equation}
and we have 
\begin{align}
&F_{mn}=x_{m}A_{n}-x_nA_m+\frac{1}{2}\eta_{mn i }\tau^i,\nonumber\\
&F_{m5}=-F_{5m}=(1+x^5)A_m.
\end{align}

On the lower patch ($x^5\le 0$), the non-compact 2nd Hopf spinor 
can be taken  as 
\begin{equation}
\psi'=\frac{1}{\sqrt{2(1-x^5)}} 
\begin{pmatrix}
(x^4+ix^i\tau_i)\phi \\
(1-x^5)\phi
\end{pmatrix},
\label{non2ndhopfspilow}
\end{equation}
and the canonical connection is  
\begin{align}
&A'_{m}=-\bar{\eta}_{mn i } \frac{x^{n}}{2(1-x^5)}\tau^i,\nonumber\\
&A'_5=0,
\label{su11connections2}
\end{align}
with  
$\bar{\eta}_{mn i }=\epsilon_{mn i 4}-\eta_{m i}\eta_{n 4}+\eta_{n i}\eta_{m 4}.$
Correspondingly, the  curvature becomes 
\begin{align}
&F'_{mn}=x_{m}A'_{n}-x_nA'_m+\frac{1}{2}\bar{\eta}_{mn i }\tau^i,\nonumber\\
&F'_{m5}=-F'_{5m}=-(1-x^5)A'_m.
\end{align}

Two expressions (\ref{explicit2ndspinor}) and (\ref{non2ndhopfspilow}) of the non-compact 2nd Hopf spinor are related by the $SU(1,1)$ transition function 
\begin{equation}
g=\frac{1}{\sqrt{1-(x^5)^2}}(x^4+ix^i\tau_i), 
\end{equation}
which satisfies $g^{\dagger}\sigma^3 g=\sigma^3$.
The transition function $g$ gives a map from $H^{2,1}$  to $SU(1,1)$, since the coordinates $\hat{x}^m=\frac{1}{\sqrt{1-(x^5)^2}}x^m$  in $g$  satisfy   $\eta_{mn}\hat{x}^m\hat{x}^n=1$.  
The differential of $g$ yields  
\begin{align}
&-ig^{\dagger}\sigma^3 dg =-\frac{1}{1-(x^5)^2}    \bar{\eta}_{mni} \sigma^3  \tau^i x^n dx^m,\nonumber\\
&-idg \sigma^3 g^{\dagger}= \frac{1}{1-(x^5)^2} \eta_{mni} \tau^i \sigma^3 x^n  dx^m.
\end{align}
Then, the canonical connections (\ref{su11connections1}) and (\ref{su11connections2}) can be rewritten as 
\begin{align}
&A_a dx^a=i\frac{1}{2}(1-x^5)dg \sigma^3 g^{\dagger}\sigma^3,\nonumber\\
&A'_a dx^a=-i\frac{1}{2} (1+x^5) \sigma^3 g^{\dagger}\sigma^3 dg,
\end{align}
and related as 
\begin{equation}
\sigma^3 A'_adx^a=g^{\dagger}(\sigma^3 A_a dx^a)g-ig^{\dagger}\sigma^3 dg.
\end{equation}
The curvatures are also 
\begin{equation}
\sigma^3 F'_{ab}=g^{\dagger}(\sigma^3 F_{ab})g.
\end{equation}

\section{3rd Non-compact Hopf Map}\label{sec3rdnoncompac}

Here, we explore a realization of the 3rd non-compact Hopf map  
\begin{equation}
H^{8,7}\overset{H^{4,3}}\longrightarrow H^{4,4}.
\end{equation}
In the constructions of the 1st and 2nd non-compact Hopf maps, the $SO(2,1)$ and $SO(3,2)$ $\it{Dirac}$ spinors are utilized, while in the 3rd non-compact Hopf map,  the $SO(5,4)$  
$\it{Majorana}$ spinor is utilized.   

\subsection{Split-Octonions}

The split-octonions, $1,e_1,e_2,\cdots,e_7$, are introduced so as to satisfy the following  relations:   
\begin{align}
&\{e_I,e_J\}=-2\eta_{IJ},\nonumber\\
&[e_I,e_J]=2f_{IJK}e_K,
\label{splitoctonionalgebra}
\end{align}
where $\eta_{IJ}=diag(+1,+1,+1,-1,-1,-1,-1)$, and $f_{IJK}$ $(I,J,K=1,2,\cdots,7)$ denotes an    antisymmetric tensor known as the structure constant of the split-octonions 
[See Table \ref{splitOctoniontable}]. 

\begin{table}
\renewcommand{\arraystretch}{1}
\hspace{1.8cm}
\begin{tabular}{|c||c|c|c|c|c|c|c|c|}
\hline       & 1  &  $e_1$ & $e_2$ & $e_3$ & $e_4$ & $e_5$ & $e_6$ & $e_7$ \\ 
\hline
\hline 1     & 1     & $e_1$   & $e_2$  & $e_3$  & $e_4$  & $e_5$  & $e_6$  & $e_7$  \\ 
\hline $e_1$ & $e_1$ & $-1$    & $e_3$  & $-e_2$ & -$e_5$  & $e_4$  & $-e_7$ & $e_6$ \\ 
\hline $e_2$ & $e_2$ & $-e_3$  & $-1$   &  $e_1$ & $-e_6$ & $e_7$  & $e_4$  & $-e_5$  \\
\hline $e_3$ & $e_3$ & $e_2$   & $-e_1$ & $-1$   & $-e_7$ & $-e_6$ & $e_5$  & $e_4$   \\ 
\hline $e_4$ & $e_4$ & $e_5$   & $e_6$  & $e_7$  & 1      & $e_1$  & $e_2$  & $e_3$ \\
\hline $e_5$ & $e_5$ & $-e_4$  & $-e_7$ & $e_6$  & $-e_1$ & 1      & $e_3$  & $-e_2$ \\
\hline $e_6$ & $e_6$ & $e_7$   & $-e_4$ & $-e_5$ & $-e_2$ & $-e_3$ & 1      & $e_1$  \\
\hline $e_7$ & $e_7$ & $-e_6$  & $e_5$  & $-e_4$ &$-e_3$  & $e_2$  & $-e_1$ & 1 \\
\hline
\end{tabular}
\caption{ The structure constants of the split-octonions can be read from the  table.  
For instance,  at the crossing point of  the column $e_1$ and the row $e_3$ we have $-e_2$:  $e_1e_3=-e_2$, which defines the corresponding structure constant as  $f_{132}=-1$. 
Similarly, we have  $f_{145}=f_{167}=f_{246}=f_{527}=f_{347}=f_{356}=-1$. Other structure constants can be derived from cyclic permutations of indices. } 
 \label{splitOctoniontable}
\end{table} 

As in the case of octonions, the split-octonions do not respect associativity as well as commutativity. 
The complex conjugation of split-octonion is defined as 
\begin{equation}
1^*=1,~~~~~(e_I)^*=-e_I.
\label{complexpropertyofocternion}
\end{equation}
Generally, with arbitrary real numbers $(r_0,r_1,r_2,\cdots,r_7)$, a split-octonion is expressed as   
\begin{equation}
o=r_0 1+\sum_{I=1}^7 r_I e_I,  
\end{equation}
and its conjugation is  
\begin{equation}
o^*=r_0 1-\sum r_I e_I.
\end{equation}
Then, we find 
\begin{equation}
o^*o=oo^*={r_0}^2+\sum_{I,J=1}^7 \eta_{IJ}{r_I}{r_J}= \sum_{I=0,1,2,3}(r_I)^2-\sum_{I=4,5,6,7}(r_I)^2,
\end{equation}
and the ``normalized'' split-octonion satisfying $o^*o=-1$ represents $H^{4,3}$. 
Similarly, the hyperbola $H_{\mathbb{O}'}^{1}$   in ``2D'' split-octonion space is given by 
\begin{equation}
o^* o-o'^* o'= \sum_{I=0,1,2,3}(r_I)^2-\sum_{I=4,5,6,7}(r_I)^2-\sum_{I=0,1,2,3}(r'_I)^2+\sum_{I=4,5,6,7}(r'_I)^2   =-1,
\end{equation}
 and $H_{\mathbb{O}'}^1$ is equivalent to $H^{8,7}$.

\subsection{Realization I}

The $SO(5,4)$ gamma matrices $\Gamma^A$ $(A=1,2,\cdots,9)$ that   satisfy 
\begin{equation}
\{\Gamma^A,\Gamma^B\}=2\eta^{AB},
\end{equation}
($\eta^{AB}=\eta_{AB}=\text{diag}(+,-,-,+,+,-,-,+,+)$) are introduced as   
\begin{align}
&\Gamma^{I}
=\begin{pmatrix}
0 & j\gamma^I \\
-j\gamma^I & 0
\end{pmatrix}, ~~\Gamma^8=
\begin{pmatrix}
0 & 1_8 \\
1_8 & 0
\end{pmatrix},~~ \Gamma^9=
\begin{pmatrix}
1_8 & 0\\
0 & -1_8
\end{pmatrix},
\end{align}
with $I=1,2,\cdots,7$. Here, $\Gamma^A$ are hermitian matrices, and  $\gamma^I$ (off-diagonal block of $\Gamma^I$) are $SO(4,3)$ gamma matrices satisfying $\{\gamma^I,\gamma^J\}=2\eta^{IJ}$, with $\eta^{IJ}=\eta_{IJ}=diag(-1,+1,+1,-1,-1,+1,+1)$. The $SO(4,3)$ gamma matrices $\gamma^I$ are explicitly  
\begin{align}
&\gamma^1=\begin{pmatrix}
0 & 0 & 0 & \sigma^2 \\
0 & 0 & -\sigma^2 & 0 \\
0 & -\sigma^2 & 0 & 0 \\
\sigma^2 & 0 & 0 & 0 
\end{pmatrix}, 
~~\gamma^2=
\begin{pmatrix}
0 & 0 & 0 & -\sigma^1 \\
0 & 0 & \sigma^1 & 0 \\
0 & \sigma^1 & 0 & 0 \\
-\sigma^1 & 0 & 0 & 0 
\end{pmatrix},\nonumber\\
&\gamma^3=
\begin{pmatrix}
0 & 0 & 0 & -\sigma^3 \\
0 & 0 & \sigma^3 & 0 \\
0 & \sigma^3 & 0 & 0 \\
-\sigma^3 & 0 & 0 & 0 
\end{pmatrix},~~
\gamma^4= j
\begin{pmatrix}
0 & 0 & 0 & 1_2\\
0 & 0 & 1_2 & 0 \\
0 & -1_2 & 0 & 0 \\
-1_2 & 0 & 0 & 0
\end{pmatrix}, \nonumber\\
&\gamma^5= j
\begin{pmatrix}
0 & 0 & 1_2 & 0 \\
0 & 0 & 0 & -1_2\\
-1_2 & 0 & 0 & 0 \\
0 & 1_2 & 0 & 0 
\end{pmatrix},~~
\gamma^6=
\begin{pmatrix}
0 & 0 & 1_2 & 0 \\
0 & 0 & 0 & 1_2 \\
1_2 & 0 & 0 & 0 \\
0 & 1_2 & 0 & 0  
\end{pmatrix},\nonumber\\
&\gamma^7=
\begin{pmatrix}
1_2 & 0 & 0 & 0 \\
0 & 1_2 & 0 & 0 \\
0 & 0 & -1_2 & 0 \\
0 & 0 & 0 & -1_2
\end{pmatrix}, 
\label{so43gammaI1}
\end{align}
where $\sigma^i$ $(i=1,2,3)$ are the split-Pauli matrices (\ref{splitimaginaryPauli}).  
The product of the purely imaginary matrices, $\gamma^1,\gamma^4,\gamma^5$,  yields a charge conjugation matrix of $SO(4,3)$: 
\begin{equation}
d=-j\gamma^1\gamma^4\gamma^5=j
\begin{pmatrix}
0 & -b \\
b & 0
\end{pmatrix},
\label{definitionofdmat}
\end{equation}
where $b$ is the $SO(3,2)$ charge conjugation matrix (\ref{chargeconjSO(3,2)}). 
Indeed, 
$d$ satisfies the relations 
\begin{equation}
-{\gamma_I}^*=d \cdot \gamma_I \cdot d^{-1},~~-\sigma_{IJ}^*=d \cdot\sigma_{IJ}\cdot d^{-1},
\end{equation}
with $SO(5,4)$ generators $\sigma_{IJ}=-j\frac{1}{4}[\gamma_I,\gamma_J]$, and has the following properties 
\begin{equation}
d^{-1}=d=d^t=d^*. 
\end{equation}
The consistency condition is satisfied as  
\begin{equation}
d^*d=dd^*(=d^2)=1_{8}.
\end{equation}
The purely imaginary matrices of $SO(5,4)$ are $\Gamma^2,\Gamma^3,\Gamma^6,\Gamma^7$, and their product yields the charge conjugation matrix of the $SO(5,4)$ group,  
\begin{equation}
B=\Gamma^2\Gamma^3\Gamma^6\Gamma^7=-
\begin{pmatrix}
d & 0 \\
0 & d
\end{pmatrix}, 
\end{equation}
which satisfies  
\begin{equation}
\Gamma_A^*=B\Gamma_A B^{-1}, ~~-\Sigma_{AB}^*=B\Sigma_{AB}B^{-1}.
\end{equation}
Here, $\Sigma_{AB}$ are the $SO(5,4)$ generators  
\begin{equation}
\Sigma_{AB}=-j\frac{1}{4}[\Gamma_A,\Gamma_B]. 
\end{equation}
The properties of $B$ are given by 
\begin{equation}
B^{-1}=B=B^t=B^*,
\end{equation}
and the consistency condition is satisfied 
\begin{equation}
B^{*}B=BB^*(=B^2)=1_{16}.
\end{equation}
With (16-component) $SO(5,4)$ spinor $\Psi$,  the Majorana condition is imposed as 
\begin{equation}
\Psi=-B^{-1}\Psi^*, 
\label{majoranacondPsi}
\end{equation}
so the $SO(5,4)$ Majorana spinor carries $16$ real degrees of freedom.
From (\ref{majoranacondPsi}),  
the $SO(5,4)$  Majorana spinor can be expressed as 
\begin{equation}
\Psi=\begin{pmatrix}
U \\
jU_c \\
V \\
jV_c
\end{pmatrix}.
\end{equation}
Here, $(U,jU_c)^t$ and $(V,jV_c)^t$ are $SO(4,3)$ 8-component Majorana spinors. 
The ``upper'' components, $U$ and $V$, denote $SO(3,2)$ 4-component Dirac spinors, and ``lower'' components, $U_c$ and $V_c$, are their charge conjugations given by Eq.(\ref{chargeconjspinorSO(3,2)}).  Then, in total, $\Psi$ carries 16 real degrees of freedom that come from $U$ and $V$. 
 
With the above preparation, we realize the 3rd non-compact Hopf map  as 
\begin{equation}
\Psi\rightarrow x^A=\Psi^{\dagger}\Gamma^A \Psi.
\label{3rdnoncompacthopf1}
\end{equation}
Here, $\Psi$ is a ``normalized'' $SO(5,4)$ Majorana spinor (3rd non-compact Hopf spinor) satisfying 
\begin{equation}
\Psi^{\dagger}\Psi=U^{\dagger}U-U_c^{\dagger}U_c+V^{\dagger}V-V_c^{\dagger}V_c=1.
\label{normalizationSO(5,4)majorana}
\end{equation}
Since $U_c^{\dagger}U_c=-U^{\dagger}U$ and $V_c^{\dagger}V_c=-V^{\dagger}V$, (\ref{normalizationSO(5,4)majorana}) can be rewritten as 
\begin{align}
&\Psi^{\dagger}\Psi=2(U^{\dagger}U+V^{\dagger}V)\nonumber\\
&~~~~~=2\sum_{\mu=1}^4
(U_R^{\mu}U_R^{\mu}+V_R^{\mu} V_R^{\mu} -U_I^{\mu}U_I^{\mu}-V_I^{\mu} V_I^{\mu})\nonumber\\
&~~~~~=1.  
\label{normalizationSO(5,4)majorana2}
\end{align}
Then, $\Psi$ denotes coordinates on $H^{8,7}$.  
The components of $x^A$ (\ref{3rdnoncompacthopf1}) are explicitly  
\begin{align}
&x^1=-2(U_1V_3+U_2V_4-U_3V_1-U_4V_2)+(c.c.), \nonumber\\
&x^2=2j(-U_1V_3+U_2V_4+U_3V_1-U_4V_2)+(c.c.),\nonumber\\
&x^3=2j(U_1V_4+U_2V_3-U_2V_2-U_4V_1)+(c.c.),\nonumber\\
&x^4=-2(U_1V_4-U_2V_3+U_3V_2-U_4V_1)+(c.c.),\nonumber\\
&x^5=-2(U_1V_2-U_2V_1-U_3V_4+U_4V_3 )+(c.c.),\nonumber\\
&x^6=2j(-U_1V_2+U_2V_1-U_3V_4+U_4V_3 )+(c.c.),\nonumber\\
&x^7= 2j(U_1^*V_1+U_2^*V_2+U_3^*V_3+U_4^*V_4) +(c.c.),\nonumber\\
&x^8=2(U_1^*V_1+U_2^*V_2+U_3^*V_3+U_4^*V_4)+(c.c.),\nonumber\\
&x^9=2(U_1^*U_1+U_2^*U_2+U_3^*U_3+U_4^*U_4)-(U\rightarrow V),
\label{explicitninexs}
\end{align}
and they satisfy  
\begin{align}
&\sum_{A,B=1,2,\cdots,9}\eta_{AB}x^A x^B\nonumber\\
&~~~~~~~=(x^1)^2- (x^2)^2 -(x^3)^2+(x^4)^2+(x^5)^2-(x^6)^2-(x^7)^2+(x^8)^2+(x^9)^2\nonumber\\
&
~~~~~~~=(\Psi^{\dagger}\Psi)^2 \nonumber\\
&~~~~~~~=1.
\end{align}
Thus, $x^A$ represent coordinates on $H^{4,4}$.  

Inverting (\ref{3rdnoncompacthopf1}) on the upper-patch of $H^{4,4}$ ($x^9\ge 0$), the  non-compact 3rd Hopf spinor is represented as  
\begin{equation}
\Psi=\frac{1}{\sqrt{2(1+x^9)}}
\begin{pmatrix}
(1+x^9) \Phi\\
(x^8+j x_I\gamma^I)\Phi
\end{pmatrix},
\label{majoranaSO(5,4)invert}
\end{equation}
where $\gamma^I$ are the $SO(4,3)$ gamma matrices (\ref{so43gammaI1}). 
From the Majorana condition of $\Psi$ (\ref{majoranacondPsi}), the 8-component spinor $\Phi$ in (\ref{majoranaSO(5,4)invert}) should satisfy 
\begin{equation}
\Phi= d \cdot \Phi^*,
\label{SO(4,3)majoranacondphi}
\end{equation}
with $d$ given by (\ref{definitionofdmat}).  
Thus, $\Phi$ denotes a $SO(4,3)$ Majorana spinor that possesses  8 real degrees of freedom,  
and $\Phi$ can be represented as 
\begin{equation}
\Phi=\frac{1}{\sqrt{2}} \begin{pmatrix}
\psi \\
j\psi_c
\end{pmatrix},
\label{PhifibreH43}
\end{equation}
where $\psi$ is a $SO(3,2)$ Dirac spinor and $\psi_c$ is its charge conjugation (\ref{chargeconjspinorSO(3,2)}). 
Moreover, from (\ref{normalizationSO(5,4)majorana2}), $\Phi$ or $\psi$ should satisfy the ``normalization condition''  
\begin{align}
\Phi^{\dagger}\Phi&=\frac{1}{2}(\psi^{\dagger}\psi-\psi_c^{\dagger}\psi_c)\nonumber\\
&= \psi^{\dagger}\psi =\sum_{\mu=1}^4\psi_R\psi_R-\sum_{\mu=1}^4\psi_I\psi_I \nonumber\\
& = 1.
\end{align}
Hence, $\psi$  denotes the  2nd non-compact Hopf spinor, and, at the same time, does 
the fibre of the 3rd non-compact Hopf map. 
Thus, we have confirmed  the hierarchical structure of the Hopf maps, and the non-compact 3rd Hopf spinor can finally be expressed as 
\begin{equation}
\Psi=\frac{1}{2\sqrt{(1+x^9)}}
\begin{pmatrix}
(1+x^9) 
{\begin{pmatrix}
\psi \\
j\psi_c
\end{pmatrix}}
\\
(x^8+j x_I\gamma^I)
{\begin{pmatrix}
\psi \\
j\psi_c
\end{pmatrix}}
\end{pmatrix}. 
\end{equation}
One might anticipate that the canonical connection of $H^{4,3}$-fibre would be induced by the formula 
\begin{equation}
A=-j\Psi^{\dagger}d\Psi.  
\end{equation}
Indeed, expressing $A$ as  
\begin{equation}
A= dx^A \Phi^{\dagger}A_A \Phi,
\label{vanishingA}
\end{equation}
we obtain 
\begin{align}
&A_{M}=-\frac{1}{1+x^9}\sigma_{MN}x^{N},\nonumber\\
&A_9=0.
\label{SO(4,4)gaugefieldsI1}
\end{align}
Here, $M,N=1,2,\cdots,8$, and $\sigma_{MN}$ are  $SO(4,4)$ ``Weyl $+$'' generators given by  
\begin{align}
&\sigma_{IJ}=-j\frac{1}{4}[\gamma_I,\gamma_J],\nonumber\\
&\sigma_{I 8}=-\sigma_{8I}=-\frac{1}{2}\gamma_I.
\label{so44weylsigma}
\end{align}
However, due to the special properties of the Majorana spinor,  the canonical connection (\ref{vanishingA}) vanishes, since antisymmetric matrix $d \cdot\sigma_{AB}$ yields $\Phi^{\dagger}\sigma_{AB}\Phi=\Phi^t (d \cdot\sigma_{AB})\Phi=0$. 
To derive the canonical connection of $H^{4,3}$-fibre, we may use   
\begin{equation}
\tilde{\Psi}=\frac{1}{\sqrt{2(1+x^9)}}
\begin{pmatrix}
(1+x^9) 1_{8}\\
(x^8 1_{8}+j x_I\gamma^I)
\end{pmatrix},
\label{matrixnoncompspi3n}
\end{equation}
and the canonical connection (\ref{SO(4,4)gaugefieldsI1}) can be derived by the formula 
\begin{equation}
A=dx^A A_A=-j\tilde{\Psi}^{\dagger}d\tilde{\Psi}. 
\end{equation}
The corresponding $SO(4,4)$ curvature  
\begin{equation}
F_{AB}=\partial_A A_B-\partial_B A_A-j[A_A,A_B]
\end{equation}
is  evaluated as 
\begin{align}
&F_{MN}=-x_{M}A_{N}+x_{N}A_{M}+\sigma_{MN},\nonumber\\
&F_{M 9}=-F_{9M}=-(1+x^9)A_{M}. 
\end{align}
On the lower patch ($x^9\le 0$), the 3rd non-compact Hopf spinor is given by 
\begin{equation}
\tilde{\Psi}'=\frac{1}{\sqrt{2(1-x^9)}}
\begin{pmatrix}
(x^8 1_{8}-j x_I\gamma^I)\\
(1-x^9) 1_{8}
\end{pmatrix},
\label{matrixnoncompspi3s}
\end{equation}
and the canonical connection is derived as 
\begin{align}
&A'_{M}=-\frac{1}{1-x^9}\bar{\sigma}_{MN}x^{N},\nonumber\\
&A'_9=0.
\label{SO(4,4)gaugefieldsI2}
\end{align}
Here, $M,N=1,2,\cdots,8$, and $\bar{\sigma}_{MN}$ are  $SO(4,4)$ ``Weyl $-$'' generators given by 
$\bar{\sigma}_{IJ}=\sigma_{IJ}$ and $\bar{\sigma}_{I 8}=-\sigma_{I8}$. 
Correspondingly, the curvature becomes  
\begin{align}
&F'_{MN}=-x_{M}A'_{N}+x_{N}A'_{M}+\bar{\sigma}_{MN},\nonumber\\
&F'_{M 9}=-F_{9M}=(1-x^9)A'_{M}. 
\end{align}

Two different expressions (\ref{matrixnoncompspi3n}) and (\ref{matrixnoncompspi3s}) of the non-compact 3rd Hopf spinor are related by the transformation 
\begin{equation}
\tilde{\Psi}'=\tilde{\Psi} \cdot 
g,
\end{equation}
where $g$ is the transition function 
\begin{equation}
g=\frac{1}{\sqrt{1-(x^9)^2}}(x^8 1_8-jx_I\gamma^I), 
\end{equation}
which satisfies $g^{\dagger}g=1$.
The coordinates $\hat{x}^M=\frac{1}{\sqrt{1-(x^9)^2}}x^M$ in $g$ satisfy the condition of  $H^{4,3}$: $\eta_{MN}\hat{x}^M\hat{x}^N=1$. Thus, $g$ represents a map from a point on  $H^{4,3}$ to a group element of $SO(4,4)$. The transition function specifies the ``gluing''  the $H^{4,3}$-fibres on upper and lower patches. 
The differential of $g$ yields 
\begin{align}
&-jg^{\dagger}dg =-\frac{2}{1-(x^9)^2}\bar{\sigma}_{MN}x^{N}dx^{M},\nonumber\\
&-jdg g^{\dagger}=\frac{2}{1-(x^9)^2}\sigma_{MN}x^{N}dx^{M}. 
\end{align}
The connections (\ref{SO(4,4)gaugefieldsI1}) and (\ref{SO(4,4)gaugefieldsI2}) can be expressed as 
\begin{align}
&A_A dx^A=j\frac{1}{2}(1-x^9)dg g^{\dagger},\nonumber\\
&A'_A dx^A=-j\frac{1}{2}(1+x^9)g^{\dagger}dg, 
\end{align}
and related as 
\begin{equation}
 A'_A dx^A=g^{\dagger} ( A_A dx^A)g-jg^{\dagger}dg. 
\end{equation}
Their curvatures are  
\begin{equation}
F'_{AB}=g^{\dagger}F_{AB} g. 
\end{equation}
Since the structure group is $SO(4,4)$, 
the canonical connection may be interpreted as the $SO(4,4)$ monopole gauge field. 

\subsection{Realization II}

As a preliminary, we first construct Majorana representation of the gamma matrices of $SO(5,4)$ group based on the split-octonion structure constants. 
With $e_0=1$, the split-octonion algebras (\ref{splitoctonionalgebra}) are expressed as  
\begin{equation}
e_I e_J=-\eta_{IJ} e_0+f_{IJK}e_K, 
\end{equation}
or 
\begin{equation}
e_A e_B=f_{ABC}e_C,
\end{equation}
where  $A,B,C=0,1,\cdots,7$. 
 (The split-octonion structure constants $f_{ABC}$ can be read from Table \ref{splitOctoniontable}.)  
With use of $f_{ABC}$, the $SO(4,3)$ gamma matrices $\lambda^I$  $(I=1,2,\cdots,7)$ are constructed as 
\begin{equation}
(\lambda^I)_{AB}=-f_{IAB},
\end{equation}
or 
\begin{align} 
&\lambda^{1}= i\left(
 \begin{array}{@{\,}cccc@{\,}}
 -\sigma^2    &   0       &     0         &   0
\\   0        & -\sigma^2 &     0         &   0
\\   0        &   0       &   \sigma^2    &   0
\\   0        &   0       &     0         & \sigma^2
 \end{array}\right),~~~ 
\lambda^{2} = \left(
 \begin{array}{@{\,}cccc@{\,}}
        0     & -\sigma^3 &      0        & 0
\\  \sigma^3   &   0       &      0        & 0
\\      0     &   0       &      0        & \sigma^3 
\\      0     &   0       &  -\sigma^3    & 0
 \end{array}
 \right),\nonumber\\
&\lambda^{3}= \left(
 \begin{array}{@{\,}cccc@{\,}}
          0   & -\sigma^1 &    0        &  0
\\  \sigma^1 &    0      &    0        &  0  
\\        0   &    0      &    0        & \sigma^1
\\        0   &    0      & -\sigma^1   &  0
 \end{array}\right),~~~~\lambda^{4} = \left(
 \begin{array}{@{\,}cccc@{\,}}
         0    &     0      &    -1_2    &   0
\\       0    &     0      &    0      &  -1_2 
\\     -1_2    &     0      &    0      &   0
\\        0   &    -1_2     &    0      &   0
 \end{array}
 \right),\nonumber\\
&\lambda^{5}= i\left(
 \begin{array}{@{\,}cccc@{\,}}
        0    &      0      &    -\sigma^2   &    0
\\      0    &      0      &      0        &  \sigma^2  
\\ \sigma^2 &      0      &      0        &    0
\\      0    &    -\sigma^2 &      0        &    0
 \end{array}\right)\!,~~~~\lambda^{6} = \left(
 \begin{array}{@{\,}cccc@{\,}}
        0     &      0      &    0        &   -1_2
\\      0     &      0      &   1_2      &    0
\\      0     &    1_2     &    0        &    0
\\     -1_2    &      0      &    0        &    0
 \end{array}
 \right),~~~~~~\nonumber\\
&\lambda^{7}= i\left(
 \begin{array}{@{\,}cccc@{\,}}
       0      &      0      &     0       &  -\sigma^2 
\\     0      &      0      &  -\sigma^2   &    0 
\\     0      & \sigma^2   &     0       &    0
\\  \sigma^2 &      0      &     0       &    0
\end{array} \right). 
\end{align}
One may check that $\lambda^I$ indeed satisfy 
\begin{equation}
\{\lambda^I,\lambda^J\}=-2\eta^{IJ}
\end{equation}
with $ \eta^{IJ}= \eta_{IJ}=\text{diag}(+1,+1,+1,-1,-1,-1,-1)$. 
 The matrices $\lambda^I$ are real matrices that satisfy  the relation 
\begin{equation}
(\lambda^I)^t=-\lambda_I.
\end{equation}
In detail, $\lambda_{1,2,3}$ are real antisymmetric matrices, and $\lambda_{4,5,6,7}$ are real symmetric matrices.  
With $\lambda^0\equiv 1_8$, $\lambda^0$ and $\lambda^I$ $(I=1,2,\cdots,7)$ represent the $SO(4,4)$ ``Weyl $+$'' gamma matrices.

With $\lambda^I$, the $SO(5,4)$ gamma matrices $\Gamma^A$  are constructed as  
\begin{equation}
\Gamma^I=i\lambda^{8-I}\otimes \sigma^2,~~\Gamma^8=1_8\otimes \sigma^1,~~\Gamma^9=1_8\otimes \sigma^3,
\end{equation}
or 
\begin{equation}
\Gamma^I=\begin{pmatrix}
0 & \lambda^{8-I} \\
-\lambda^{8-I} & 0
\end{pmatrix},~~
\Gamma^8=
\begin{pmatrix}
0 & 1_8 \\
1_8 & 0 
\end{pmatrix},~~
\Gamma^9=
\begin{pmatrix}
1_8 & 0 \\
0 & -1_8 
\end{pmatrix}.
\end{equation}
Thus, the split-octonion structure constants appear in 
the off-diagonal elements of $\Gamma^A$.  
The matrices $\Gamma^A$ satisfy  
\begin{equation}
\{\Gamma^A,\Gamma^B\}=2\eta^{AB},
\end{equation}
with $\eta^{AB}=\eta_{AB}=\text{diag}(-1,-1,-1,-1,+1,+1,+1,+1,+1)$.  
They are non-hermitian:  
\begin{equation}
(\Gamma^A)^t=\Gamma_A.
\end{equation}
In detail, $\Gamma^{1,2,3,4}$ are antisymmetric real matrices, and $\Gamma^{5,6,7,8,9}$ are  symmetric real matrices.  
The $SO(5,4)$ generators are constructed as  
\begin{equation}
\Sigma_{AB}=-i\frac{1}{4}[\Gamma_A,\Gamma_B],
\label{IIso54generat}
\end{equation}
or 
\begin{align}
&\Sigma_{IJ}=\begin{pmatrix}
\sigma_{IJ} & 0 \\
0 & \sigma_{IJ}
\end{pmatrix},~~~\Sigma_{I8}=-\frac{i}{2}\begin{pmatrix}
\lambda_{8-I} & 0 \\
0 & -\lambda_{8-I}
\end{pmatrix},\nonumber\\
&\Sigma_{I9}= \frac{i}{2} 
\begin{pmatrix}
0 & \lambda_{8-I}  \\
\lambda_{8-I} & 0
\end{pmatrix},~~~\Sigma_{89}=-i\frac{1}{2}
\begin{pmatrix}
0 & -1_8 \\
1_8 & 0
\end{pmatrix},
\end{align}
where $\sigma_{IJ}$ are the $SO(3,4)$ generators given by 
\begin{equation}
\sigma_{IJ}=i\frac{1}{4}[\lambda_{8-I},\lambda_{8-J}].
\end{equation}
Since $\Gamma^A$ are real matrices, the corresponding $SO(5,4)$ generators (\ref{IIso54generat}) are purely imaginary and satisfy the relation; $\Sigma_{AB}^*=-\Sigma_{AB}$. Thus,  the present representation is indeed the Majorana representation, in which the charge conjugation matrix is given by unit matrix. 

Although not all of $\Gamma^{A}$ are hermitian, we can construct  symmetric real matrices as 
\begin{equation}
K^A=K\Gamma^A,
\end{equation}
where 
\begin{equation}
K=
\begin{pmatrix}
\Sigma^3 & 0 \\
0 & \Sigma^3 
\end{pmatrix},
\end{equation}
with 
\begin{equation}
\Sigma^3=
\begin{pmatrix}
1_4 & 0 \\
0 & -1_4
\end{pmatrix}. 
\end{equation}
The symmetric real matrices  $K^A$ are explicitly given by 
\begin{align}
&K^I=
\begin{pmatrix}
0 & \Sigma^3 \lambda^{8-I} \\
-\Sigma^3 \lambda^{8-I} & 0
\end{pmatrix},~~
K^8=
\begin{pmatrix}
0 & \Sigma^3\\
\Sigma^3 & 0 
\end{pmatrix},~~\nonumber\\
&K^9=
\begin{pmatrix}
\Sigma^3 & 0 \\
0 & -\Sigma^3
\end{pmatrix}, 
\end{align}
where $\Sigma^3\lambda^I$ are the following real antisymmetric matrices:   
\begin{align}
&\Sigma^3\lambda^1=-i
\begin{pmatrix}
\sigma^2 & 0 & 0 & 0\\
0 & \sigma^2 & 0 & 0 \\
0 & 0 & \sigma^2 & 0\\
0 & 0 & 0 & \sigma^2
\end{pmatrix},~~
\Sigma^3\lambda^2=
\begin{pmatrix}
0 & -\sigma^3 & 0 & 0 \\
\sigma^3 & 0 & 0 & 0 \\
0 & 0 & 0 & -\sigma^3 \\
0 & 0 & \sigma^3 & 0
\end{pmatrix},\nonumber\\
&\Sigma^3 \lambda^3=
\begin{pmatrix}
0 & -\sigma^1 & 0 & 0 \\
\sigma^1 & 0 & 0 & 0 \\
0 & 0 & 0 & -\sigma^1 \\
0 & 0 & \sigma^1 & 0 
\end{pmatrix},~~
\Sigma^3\lambda^4
=\begin{pmatrix}
0 & 0 & 1_2 & 0 \\
0 & 0 & 0 & 1_2\\
-1_2 & 0 & 0 & 0 \\
0 & -1_2 & 0 & 0 
\end{pmatrix},\nonumber\\
&\!\!\Sigma^3\lambda^5=i
\begin{pmatrix}
0 & 0 & \sigma^2 & 0  \\
0 & 0 & 0 & -\sigma^2 \\
\sigma^2 & 0 & 0 & 0 \\
0 & -\sigma^2 & 0 & 0 
\end{pmatrix}, ~~
\Sigma^3\lambda^6=
\begin{pmatrix}
0 & 0 & 0 & 1_2\\
0 & 0 & -1_2 & 0\\
0 & 1_2 & 0 & 0 \\
-1_2 & 0 & 0 & 0
\end{pmatrix},\nonumber\\
&\Sigma^3 \lambda^7 =i
\begin{pmatrix}
0 & 0 & 0 & \sigma^2\\
0 & 0 & \sigma^2 & 0 \\
0 & \sigma^2 & 0 & 0 \\
\sigma^2 & 0 & 0  & 0
\end{pmatrix}.
\end{align}
In the Majorana representation, the $SO(5,4)$ Majorana spinor 
is simply given by the (16-component) real spinor 
\begin{equation}
\Psi^*=\Psi. 
\end{equation}

We now realize the 3rd non-compact Hopf map as 
\begin{equation}
\Psi\rightarrow x^A=\Psi^t K^A \Psi,
\label{map3rdhopfII}
\end{equation}
where $\Psi$ is a ``normalized'' $SO(5,4)$ Majorana spinor (the  non-compact 3rd Hopf spinor) satisfying 
\begin{equation}
\Psi^t K\Psi =\sum_{\mu=1}^4 {\Psi_{\mu}}^2 +\sum_{\mu=9}^{12}{\Psi_{\mu}}^2-\sum_{\mu=5}^8{\Psi_{\mu}}^2 -\sum_{\mu=13}^{16}{\Psi_{\mu}}^2 =1, 
\end{equation}
which represents $H^{8,7}$. 
With two $SO(4,4)$ Majorana-Weyl spinors, $\Psi^{+}$ and $\Psi^{-}$ (each of which represents 8-component real spinor), the $SO(5,4)$ Majorana spinor $\Psi$ can be expressed as 
\begin{equation}
\Psi=
\begin{pmatrix}
\Psi^{+}\\
\Psi^{-}
\end{pmatrix},
\end{equation}
and the components of  $x^A$ (\ref{map3rdhopfII}) are given by 
\begin{align}
&x^I=2{(\Psi^+)}^t\Sigma^3\lambda^{8-I} \Psi^-,\nonumber\\
&x^8=2{(\Psi^+)}^t \Sigma^3\Psi^-,\nonumber\\
&x^9={(\Psi^+)}^t \Sigma^3 \Psi^+ -{(\Psi^-)}^t \Sigma^3\Psi^-,
\end{align}
or more explicitly 
\begin{align}
&x^1= \Psi^+_1 \Psi^-_8-\Psi^+_2 \Psi^-_7+\Psi^+_3 \Psi^-_6-\Psi^+_4 \Psi^-_5-(\Psi^+\leftrightarrow \Psi^-), \nonumber\\
&x^2= \Psi^+_1 \Psi^-_7+\Psi^+_2 \Psi^-_8-\Psi^+_3 \Psi^-_5-\Psi^+_4 \Psi^-_6-(\Psi^+\leftrightarrow \Psi^-), \nonumber\\
&x^3=  \Psi^+_1 \Psi^-_6-\Psi^+_2 \Psi^-_5-\Psi^+_3 \Psi^-_8+\Psi^+_4 \Psi^-_7-(\Psi^+\leftrightarrow \Psi^-), \nonumber\\
&x^4= \Psi^+_1 \Psi^-_5+\Psi^+_2 \Psi^-_6+\Psi^+_3 \Psi^-_7+\Psi^+_4 \Psi^-_8-(\Psi^+\leftrightarrow \Psi^-), \nonumber\\
&x^5= -\Psi^+_1 \Psi^-_4-\Psi^+_2 \Psi^-_3 -\Psi^+_5 \Psi^-_8 -\Psi^+_6 \Psi^-_7-(\Psi^+\leftrightarrow \Psi^-), \nonumber\\
&x^6= -\Psi^+_1 \Psi^-_3+\Psi^+_2 \Psi^-_4 -\Psi^+_5 \Psi^-_7 + \Psi^+_6 \Psi^-_8 -(\Psi^+\leftrightarrow \Psi^-), \nonumber\\
&x^7= -\Psi^+_1 \Psi^-_2-\Psi^+_3\Psi^-_4-\Psi^+_5\Psi^-_6-\Psi^+_7\Psi^-_8-(\Psi^+\leftrightarrow \Psi^-), \nonumber\\
&x^8= \Psi^+_{1} \Psi^-_{1} +\Psi^+_{2} \Psi^-_{2} +\Psi^+_{3} \Psi^-_{3} +\Psi^+_{4} \Psi^-_{4} \nonumber\\
&~~~~  -\Psi^+_{5} \Psi^-_{5} -\Psi^+_{6} \Psi^-_{6}-\Psi^+_{7} \Psi^-_{7}-\Psi^+_{8} \Psi^-_{8},\nonumber\\
&x^9= ({\Psi^+_1}^2+{\Psi^+_2}^2+{\Psi^+_3}^2+{\Psi^+_4}^2) -({\Psi^+_5}^2+{\Psi^+_6}^2+{\Psi^+_7}^2+{\Psi^+_8}^2)\nonumber\\
&~~~~ -({\Psi^-_1}^2+{\Psi^-_2}^2+{\Psi^-_3}^2+{\Psi^-_4}^2) +({\Psi^-_5}^2+{\Psi^-_6}^2+{\Psi^-_7}^2+{\Psi^-_8}^2).  
\end{align}
A straightforward calculation shows that $x^A$ satisfy 
\begin{align}
&\sum_{A,B=1,2,\cdots,9}\eta_{AB}x^A x^B\nonumber\\
&~~~=-(x^1)^2-(x^2)^2-(x^3)^2-(x^4)^2 +(x^5)^2+(x^6)^2+(x^7)^2+(x^8)^2+(x^9)^2
\nonumber\\
&~~~=( \sum_{\mu=1}^4 (\Psi_{\mu}^{+})^2 -\sum_{\mu=5}^8 (\Psi_{\mu}^{+})^2 +\sum_{\mu=1}^4 (\Psi_{\mu}^{-})^2 -\sum_{\mu=5}^8 (\Psi_{\mu}^{-})^2 )^2 \nonumber\\
&~~~=(\Psi^t K\Psi)^2\nonumber\\
&~~~=1,
\end{align}
which suggests that $x^A$ are regarded as coordinates on $H^{4,4}$.  
Inverting (\ref{map3rdhopfII}) on the upper patch of $H^{4,4}$ ($x^9\ge 0$), $\Psi$ can be represented as  
\begin{equation}
\Psi= \frac{1}{\sqrt{2(1+x^9)}}
\begin{pmatrix}
(1+x^9) 
\Phi \\
(x^8-\lambda_{8-I} x^I) \Phi
\end{pmatrix}, 
\end{equation}
where $\Phi$ is the $SO(4,3)$ real 8-component spinor subject to the constraint 
\begin{equation}
\Phi^t \Sigma^3\Phi= \sum_{\mu=1}^4 (\Phi_{\mu})^2-\sum_{\mu=5}^8(\Phi_{\mu})^2      =1, 
\end{equation}
and then, $\Phi$ denotes $H^{4,3}$-fibre. 
As discussed below Eq.(\ref{so44weylsigma}), naively anticipated canonical connection $A=-i\Psi^{t} K d\Psi$ vanishes due to the Majorana property of $\Psi$. 
Hence, defining 
\begin{equation}
\tilde{\Psi}=
\frac{1}{\sqrt{2(1+x^9)}}
\begin{pmatrix}
(1+x^9) 1_{8} \\
x^8 1_8-\lambda_{8-I} x^I 
\end{pmatrix}, 
\label{nc3dhopf1}
\end{equation}
we evaluate the canonical connection of $H^{4,3}$-fibre 
\begin{equation}
A=-i\tilde{\Psi}^t K d\tilde{\Psi}=dx^A \Sigma^3 A_A, 
\end{equation}
to derive 
\begin{align}
&A_{M}=\frac{1}{1+x^9}\sigma_{MN}x^{N},\nonumber\\
&A_9=0.
\label{2conneso44I}
\end{align}
Here, $\sigma_{MN}$ are $SO(4,4)$ ``Weyl +'' generators  given by 
\begin{equation}
\sigma_{IJ}=-i\frac{1}{4}[\lambda_I,\lambda_J],~~~~\sigma_{I 8}=-\sigma_{8 I}=i\frac{1}{2}\lambda_I.
\end{equation}
($\Sigma^3\sigma_{IJ}$ and $\Sigma^3\sigma_{I 8}$ are pure imaginary antisymmetric matrices.) 
Similarly, the corresponding curvature  
\begin{equation}
F_{AB}=\partial_A A_B-\partial_B A_A -i[A_A,A_B]
\end{equation}
is computed as 
\begin{align}
&F_{MN}=-x_{M} A_{N}+x_{N}A_{M}-\sigma_{MN},\nonumber\\
&F_{M 9}=-F_{9M}=(1+x^9)A_{M}.
\end{align}

In the lower patch ($x^9 \le 0$), the 3rd non-compact Hopf spinor is given by 
\begin{equation}
\tilde{\Psi}'=
\frac{1}{\sqrt{2(1-x^9)}}
\begin{pmatrix}
x^8 1_8+\lambda_{8-I} x^I \\
(1-x^9) 1_{8} 
\end{pmatrix}, 
\label{nc3dhopf2}
\end{equation}
and the canonical connection is derived as 
\begin{align}
&A'_{M}=\frac{1}{1-x^9}\bar{\sigma}_{MN}x^{N},\nonumber\\
&A'_9=0,
\label{2conneso44II}
\end{align}
where $\bar{\sigma}_{IJ}=\sigma_{IJ}$, and $\bar{\sigma}_{I8}=-\sigma_{I8}$. 
Correspondingly, the curvature becomes  
\begin{align}
&F'_{MN}=-x_{M} A'_{N}+x_{N}A'_{M}-\bar{\sigma}_{MN},\nonumber\\
&F'_{M 9}=-F'_{9M}=-(1-x^9)A'_{M}.
\end{align}

The transition function that relates two different expressions (\ref{nc3dhopf1}) and (\ref{nc3dhopf2}) of the 3rd non-compact spinor is given by  
\begin{equation}
g=\frac{1}{\sqrt{1-(x^9)^2}}(x^8 1_8 +\lambda_{8-I}x^I),
\end{equation}
which satisfies $g^{\dagger}\Sigma^3 g=\Sigma^3$. 
The transition function $g$ is a function from $H^{4,3}$ to $SO(4,4)$, and 
its differential yields  
\begin{align}
&-ig^{\dagger}\Sigma^3 dg=\frac{2}{1-(x^9)^2}\Sigma^3 \bar{\sigma}_{MN}x^Ndx^M,\nonumber\\
&-idg \Sigma^3 g^{\dagger}=-\frac{2}{1-(x^9)^2}\sigma_{MN} \Sigma^3 x^N dx^M.
\end{align}
Then, the canonical connections (\ref{2conneso44I}) and (\ref{2conneso44II}) can be expressed as 
\begin{align}
&A_A dx^A=i\frac{1-x^9}{2} dg \Sigma^3 g^{\dagger}\Sigma^3,\nonumber\\
&A'_A dx^A =-i\frac{1+x^9}{2}\Sigma^3 g^{\dagger}\Sigma^3 dg,
\end{align}
and they are related as 
\begin{equation}
\Sigma^3 A'_A dx^A=g^{\dagger}(\Sigma^3 A_A dx^A) g-ig^{\dagger}\Sigma^3 dg.
\end{equation}
Their curvatures are also  
\begin{equation}
\Sigma^3 F'_{AB}=g^{\dagger}(\Sigma^3 F_{AB}) g.
\end{equation}

\section{Summary and Discussion}\label{secsummary}

 Based on the split-algebras, we developed the non-compact version of the Hopf maps. 
Simply replacing the ordinary imaginary unit with the split-imaginary unit, the ultra-hyperboloids with split-signatures were naturally introduced, and the non-compact Hopf maps were  straightforwardly constructed.  
With the explicit realization of the non-compact Hopf maps, the topological structures of the associated non-compact fibres were explored. 

As briefly mentioned in Introduction, the original Hopf fibrations correspond to the fibre configurations of the $U(1)$, $SU(2)$ and $SO(8)$ monopoles with minimum charge. The stability of such  monopoles with arbitrary charges are topologically accounted by the homotopy theorems 
\begin{equation}
\pi_1 (U(1)) \simeq \mathbb{Z},
~~~\pi_3 (SU(2)) \simeq \mathbb{Z},~~~~\pi_7 (SO(8)) \simeq \mathbb{Z}\oplus \mathbb{Z}.
\end{equation}
 Transition function plays a crucial role in specifying non-trivial realization of topology of fibre-bundle. For instance, in the $SU(2)$ monopole case, the transition function represents a map from the $S^3$-equator of base manifold $S^4$ to the $SU(2)$ structure group.  Then, the homotopy  theorem that accounts for the topological stability of the $SU(2)$ monopole  becomes  $\pi_3(SU(2))\simeq \mathbb{Z}$. 
Meanwhile, in the present, each of the three non-compact Hopf maps is expected to correspond to the fibre configuration of $SO(1,1)$, $SU(1,1)$, and $SO(4,4)$ monopoles with minimum charge. For instance, in the case of the 2nd non-compact Hopf map,  the transition function represents a map from the $H^{2,1}$-``equator''  to the structure group $SU(1,1)$.  The group manifold of $SU(1,1)$ is topologically equivalent to $H^{2,1}$, so the transition function represents a map between identical spaces. However, $H^{2,1}$ is non-compact, and   
the winding between the non-compact spaces has to be concerned.  The homotopy theorems between non-compact spaces are required to guarantee topological stability of  non-compact monopoles.  The homotopy theorem for non-compact monopoles has to be exploited in a  future research.  

As also mentioned in Introduction, there are wide applications of the original Hopf maps to physics. 
In particular, there are intriguing relations reported between the division algebras and the $\mathcal{N}=1$ supersymmetric Yang-Mills theory and Green-Schwarz superstring formalism in special dimensions \cite{KugoTownsend1983,Evans1988}. 
In the language of group theory, the 1st Hopf map is related to the $SL(2,\mathbb{C})$ group which is the (3+1)D Lorentz group, and the octonion group related to the 3rd Hopf map  corresponds to the Lorentz group in $(9+1)$D \cite{sudbery1984,chungsudbery1987}: the critical dimension of the superstring theory. 
In the case of the non-compact Hopf versions, one may infer similar correspondences (see Table \ref{lorentzgroupsandalgebras}), and the split-algebras naturally bring the notion of space-time with split-signatures (split-space-time) \cite{FootJoshi1990,FootJoshi1992}. 
 It is stimulating, and may be worthwhile, to speculate  
 applications of the split-algebras to physics.   
Indeed, recently,  analogies of quantum Hall effect and the twistor theory were explored based on the 2nd non-compact  Hopf map \cite{arXiv:0902.2523}. 
Although direct relations to the split-algebras have not been discussed yet,  
split-space-time groups play a crucial role in the  ``doubled geometry''  formulation of string theory \cite{hep-th/0605149}.   
Since split-space-times generally have more than one-time dimension, one may be anxious about  
   the negative-norm problem. However, at the same time, it may also be probable that dualities between space-times with different signature metrics secure the theories from the fatal negative norm problems \cite{hep-th/9808069,hep-th/0610122}. 
Apparently, this also needs more investigations.

\begin{table}
\renewcommand{\arraystretch}{1}
\hspace{5mm}
\begin{tabular}{|c|c|c|}
\hline    &   Division algebras & ~~~~~~ Split algebras ~~~~ 
\\
\hline\hline  Real numbers  &   $SO(2,1) \simeq SL(2,\mathbb{R})$ 
&  $SO(2,1) \simeq SL(2,\mathbb{R})$  \\ 
\hline Complex numbers &  $SO(3,1)\simeq SL(2,\mathbb{C})$ & $SO(2,2)\simeq SL(2,\mathbb{C}')$ \\
\hline
Quaternions &    $SO(5,1)\simeq SL(2,\mathbb{H})$ &    $SO(3,3)\simeq SL(2,\mathbb{H}')$\\
 \hline
Octonions &  $SO(9,1)\simeq SL(2,\mathbb{O})$ & $SO(5,5)\simeq SL(2,\mathbb{O}')$\\
\hline
\end{tabular}
\caption{Lorentz groups and split-Lorentz groups} 
 \label{lorentzgroupsandalgebras}
\end{table} 

Several other generalizations of the Hopf maps have also been reported. 
For instance, the supersymmetric version of the 1st Hopf map was explored in Refs.\cite{LandiMarmo1987,math-ph/9907020,hep-th/0409230}. (We have given the non-compact supersymmetric Hopf map \cite{arXiv:0809.4885} in Appendix.) 
Although yet to be found, it might even exist supersymmetric versions of the 
2nd and 3rd Hopf maps \footnote{An attempt to this direction, one may see Ref.\cite{arXiv:0902.2682}.}.
The $\theta$-deformed 2nd Hopf map has also been explored in Ref.\cite{LandiPaganiReinamath/0407342}. 
Such generalization is also applicable to the present non-compact Hopf maps. 

 Finally, we comment on a relation to the split-instanton configuration introduced by Mason \cite{math-ph/0505039}. The split-instanton is a solution of the anti-self-dual equation of Yang-Mills fields on ${R}^{2,2}$. The $SU(1,1)$ monopole is related to such split-instanton by conformal transformation between ${H}^{2,2}$ and ${R}^{2,2}$. This is  hyperbolic analogue of the  conformal relation between the $SU(2)$ monopole field configuration on $S^4$ and the $SU(2)$ instanton configuration on ${R}^4$.

\section*{Note Added}

After completion of this work, the author learned the work of Bla$\check{\text{z}}$i$\acute{\text{c}}$ \cite{Blazic1996} where coset construction of para-quaternionic projective spaces was discussed. The present construction of the non-compact 2nd Hopf map is consistent with the work. 

\section*{Acknowledgement}
The author would like to thank Professor Mikio Nakahara for useful conversations. 
This work was partially supported by Sumitomo foundation. 

\appendix

\section*{Appendix}

\section{Non-compact Supersymmetric Hopf Map}\label{secsusynon}

A supersymmetric generalization of the (1st) Hopf map is introduced by replacing the original bosonic manifolds with  their supersymmetric counterparts \cite{LandiMarmo1987,math-ph/9907020,hep-th/0409230}: 
\begin{equation}
S^{3|2}\overset{S^1} \longrightarrow S^{2|2}.
\label{compactsusyhopf}
\end{equation}
The bosonic body of the base-manifold $S^{2|2}$ is $S^2$, and the isometry group of $S^{2|2}$ is $UOSp(1|2)$  with  bosonic subgroup  $SU(2)$. 

Meanwhile, the non-compact version of the supersymmetric Hopf map is given by   
\begin{equation}
H^{2,1|2}\overset{H^{1,0}}\longrightarrow H^{1,1|2},
\label{susy1sthopfmap}
\end{equation}
or 
\begin{equation}
AdS^{3|2} \overset{ \mathcal{U}(1)}\longrightarrow AdS^{2|2}.
\end{equation}
Here, the body of the base-manifold $H^{1,1|2}$ is $H^{1,1}$, and 
the isometry group of $AdS^{2|2}$ is $OSp(1|2)$ 
with bosonic subgroup $SU(1,1)$. 

\subsection{Realization I}

The $OSp(1|2)$ algebra is given by  
\begin{align}
&[l^i,l^j]=j\epsilon^{ijk}l_k,\nonumber\\
&[l^i,l^{\alpha}]=\frac{1}{2}(\sigma^i)_{\beta}^{~~\alpha}l^{\beta},\nonumber\\
&\{l^{\alpha},l^{\beta}\}=\frac{1}{2}(\epsilon \sigma^i)^{\alpha\beta}l_i,
\end{align}
where $\epsilon^{123}=1$, and $\eta_{ij}=\text{diag}(+1,-1,+1)$. 
With the split-Pauli matrices $\sigma^i$, and $\tau^{1}=(1,0)^t$, $\tau^2=(0,1)^t$, the  
 $OSp(1|2)$ generators are  represented as 
\begin{equation}
l^i=\frac{1}{2}
\begin{pmatrix}
\sigma^i & 0 \\
0 & 0
\end{pmatrix},~~~
l^{\alpha}=
\frac{1}{2}
\begin{pmatrix}
0 & \tau^{\alpha}\\
-(\epsilon \tau^{\alpha})^t & 0 
\end{pmatrix},
\end{equation}
or
\begin{align}
&l^1=\frac{1}{2}
\begin{pmatrix}
0 & 1 & 0 \\
1 & 0 & 0 \\
0 & 0 & 0 
\end{pmatrix},~~~
l^2=\frac{1}{2}
\begin{pmatrix}
0 & -j & 0 \\
j & 0 & 0 \\
0 & 0 & 0 
\end{pmatrix},~~~
l^3=\frac{1}{2}
\begin{pmatrix}
1 & 0 & 0 \\
0 & -1 & 0 \\
0 & 0 & 0 
\end{pmatrix},\nonumber\\
&l^{\theta_1}=\frac{1}{2} 
\begin{pmatrix}
 0 & 0 & 1 \\
 0 & 0 & 0\\
 0 & 1 & 0
\end{pmatrix},~~
l^{\theta_2}=\frac{1}{2} 
\begin{pmatrix}
 0 & 0 & 0 \\
 0 & 0 & 1\\
 -1 & 0 & 0
\end{pmatrix}.
\end{align}
The $OSp(1|2)$ spinor takes the form of 
\begin{equation}
\chi=
\begin{pmatrix}
u \\
v\\
\eta
\end{pmatrix},
\end{equation}
where the first two components $u$ and $v$ are Grassmann even, while the last component $\eta$ is Grassmann odd. 

With $l^i$ and $l^{\alpha}$, the non-compact supersymmetric Hopf map (\ref{susy1sthopfmap}) is realized as 
\begin{equation}
\chi ~~\rightarrow~~ x^i=2\chi^{\ddagger}l^i\chi,~~\theta^{\alpha}=2\chi^{\ddagger}l^{\alpha}\chi.
\label{superhopf1}
\end{equation}
Here, $\chi$ is a  $OSp(1|2)$ spinor subject to the ``normalization condition'' 
\begin{equation}
\chi^{\ddagger}\chi=u^*u+v^*v-\eta^*\eta=(u_R)^2+(v_R)^2-(u_I)^2-(v_I)^2-\eta^*\eta=1,
\end{equation}
where the superadjoint $\ddagger$ is defined as  
\footnote{The symbol $*$ represents the pseudo-conjugation, $(\eta^*)^*=-\eta$, $(\eta_1\eta_2)^*=\eta_1^*\eta_2^*$ for the Grassmann odd quantities $\eta_1$ and $\eta_2$. In particular, $(\theta^1)^*=\theta^2$,  $(\theta^2)^*=-\theta^1$. }
\begin{equation}
\chi^{\ddagger}=(u^*,v^*,-\eta^*).
\end{equation}
The components of  $x^i$ and $\theta^{\alpha}$ (\ref{superhopf1}) are explicitly 
\begin{align}
&x^1=u^*v+v^*u,~~x^2=-j u^*v+jv^*u,~~x^3=u^*u-v^*v,\nonumber\\
&\theta^{1}=u^*\eta-\eta^* v,~~\theta^2=v^*\eta+\eta^*u, 
\end{align}
and  satisfy 
\begin{equation}
\eta_{ij}x^ix^j+\epsilon_{\alpha\beta}\theta^{\alpha}\theta^{\beta}=1.
\end{equation}
Thus, $x^i$ and $\theta^{\alpha}$ denote coordinates on  $H^{1,1|2}$. 
On the upper patch of $H^{1,1|2}$ ($x^3\ge 0$), inverting the map (\ref{superhopf1}),  $\chi$ can be expressed as 
\begin{equation}
\chi=\frac{1}{\sqrt{2(1+x^3)}}
\begin{pmatrix}
(1+x^3) (1-\frac{1}{4(1+x^3)}\theta\epsilon\theta)\\
(x^1-jx^2)(1+\frac{1}{4(1+x^3)}\theta\epsilon\theta)\\
(1+x^3)\theta^1+(x^1-jx^2)\theta^2
\label{2chiupper}
\end{pmatrix}.
\end{equation}
The canonical connection of $\mathcal{U}(1)$-fibre is induced as 
\begin{equation}
A=-j\chi^{\ddagger}d\chi=dx^i A_i+d\theta^{\alpha}A_{\alpha},
\end{equation}
with 
\begin{align}
&A_i=\frac{1}{2(1+x^3)}\epsilon_{ij3}x^j\biggl(1+\frac{2+x^3}{2(1+x^3)}\theta\epsilon\theta\biggr),\nonumber\\
&A_{\alpha}=\frac{1}{2}j(x_i \sigma^i\epsilon \theta)_{\alpha}.
\label{1superconneI}
\end{align}
The curvature is given by the formula 
\begin{equation}
F=dA =\frac{1}{2}dx^i \wedge dx^j F_{ij}+dx^i \wedge d\theta^{\alpha} F_{i\alpha}-\frac{1}{2}d\theta^{\alpha}\wedge d\theta^{\beta} F_{\alpha\beta}, 
\end{equation}
where 
\begin{equation}
F_{ij}=\partial_i A_j-\partial_j A_i, ~~~F_{i\alpha}=\partial_i A_{\alpha}-\partial_{\alpha}A_i, ~~~F_{\alpha\beta}=\partial_{\alpha}A_{\beta}+\partial_{\beta}A_{\alpha}.
\end{equation}
They are evaluated as  
\begin{align}
&F_{ij}=-\frac{1}{2}\epsilon_{ij3}x^j\biggl(1+\frac{3}{2}\theta\epsilon\theta\biggr),\nonumber\\
&F_{i\alpha}=\frac{1}{2}j(\sigma^j\epsilon\theta)_{\alpha}(\eta_{ij}-3x_ix_j),\nonumber\\
&F_{\alpha\beta}=jx^i(\sigma_i\epsilon)_{\alpha\beta}\biggl(1+\frac{3}{2}\theta\epsilon\theta\biggr).
\end{align}

On the lower patch ($x^3\le 0$), the ``normalized'' $OSp(1|2)$ Hopf spinor can similarly be expressed as 
\begin{equation}
\chi'=\frac{1}{\sqrt{2(1-x^3)}}
\begin{pmatrix}
(x^1+jx^2)(1+\frac{1}{4(1-x^3)}\theta\epsilon\theta)\\
(1-x^3) (1-\frac{1}{4(1-x^3)}\theta\epsilon\theta)\\
(x^1+jx^2)\theta^1+(1-x^3)\theta^2 
\label{2chilow}
\end{pmatrix}, 
\end{equation}
and the corresponding canonical connection is derived as 
\begin{align}
&A'_i=-\frac{1}{2(1-x^3)}\epsilon_{ij3}x^j\biggl(1+\frac{2-x^3}{2(1-x^3)}\theta\epsilon\theta\biggr),\nonumber\\
&A'_{\alpha}=\frac{1}{2}j(x_i \sigma^i\epsilon \theta)_{\alpha}=A_{\alpha}.
\label{1superconneII}
\end{align}
Two expressions (\ref{2chiupper}) and (\ref{2chilow}) are related by the transformation  
\begin{equation}
\chi'=\chi g,
\end{equation}
where $g$ is the transition function of the form   
\begin{equation}
g=\frac{x^1+jx^2}{\sqrt{1-(x^3)^2}}\biggl(1+\frac{1}{2(1-(x^3)^2)}\theta\epsilon\theta\biggr), 
\end{equation}
which satisfies $g^*g=1$ and represents a map from $H^{1,0}$ to $\mathcal{U}(1)$. 
The differential of $g$ yields 
\begin{equation}
-jg^*dg =-jdg g^*=-\frac{1}{1-(x^3)^2}\epsilon_{ij3}x^j \biggl(1+\frac{1}{1-(x^3)^2}\theta\epsilon\theta\biggr)dx^i, 
\end{equation}
and the canonical connections (\ref{1superconneI}) and (\ref{1superconneII}) are expressed as 
\begin{align}
&A=-j\biggl(-\frac{1-x^3}{2}+\frac{x^3}{4}\theta\epsilon\theta\biggr)g^{*}dg+d\theta^{\alpha}A_{\alpha},\nonumber\\
&A'=-j\biggl(\frac{1+x^3}{2}+\frac{x^3}{4}\theta\epsilon\theta\biggr)g^{*}dg+d\theta^{\alpha}A_{\alpha},
\end{align}
and related 
\begin{equation}
A'=A-jg^{*}dg. 
\end{equation}
Similarly, their curvatures are  
\begin{equation}
F'=F.
\end{equation}
The present canonical connection may be interpreted as the (non-compact) gauge field of supermonopole. 

\subsection{Realization II}

As in the original bosonic case, in Realization II, the $\mathcal{U}(1)$ structure group is replaced by the ordinary $U(1)$ group, and the non-compact supersymmetric Hopf map (\ref{susy1sthopfmap}) 
 is  modified  as  \cite{arXiv:0809.4885} 
\begin{equation}
 H^{2,1|2} \overset{H^{0,1}}\longrightarrow H^{2,0|2},
\label{superhopfII}
\end{equation}
or 
\begin{equation}
AdS^{3|2}\overset{U(1)}\longrightarrow \text{Eucl.} AdS^{2|2}. 
\end{equation}
The $OSp(1|2)$ algebra is given by 
\begin{align}
&[l^i,l^j]=i\epsilon^{ij}_{~~k}l^k,\nonumber\\
&[l^i,l^{\alpha}]=\frac{1}{2}(\tau^i)_{\beta}^{~~\alpha}l^{\beta},\nonumber\\
&\{l^{\alpha},l^{\beta}\}=\frac{1}{2}(\epsilon^t \tau_i)^{\alpha\beta}l^i,
\label{ospr12algebra}
\end{align}
where $\tau^i$ are the $SU(1,1)$ generators (\ref{defoftaus}), and  $\eta_{ij}=diag(+1,+1,-1)$. 
The $OSp(1|2)$ generators are 
\begin{equation}
l^i=\frac{1}{2}
\begin{pmatrix}
\tau^i & 0 \\
0 & 0
\end{pmatrix},
~~l^{\alpha}=
\frac{1}{2}
\begin{pmatrix}
0 & \tau^{\alpha}\\
-(\epsilon \tau^{\alpha})^t & 0 
\end{pmatrix}.
\label{originalospfundmatr}
\end{equation}
The complex representation is given by 
$\tilde{l}^i=-(l^i)^*$ and $\tilde{l}^{\alpha}=\epsilon_{\alpha\beta}l^{\beta}$, and they are related to the original representation as 
\begin{equation}
\tilde{l}^i=\mathcal{R}^{\dagger}l^i\mathcal{R},~~\tilde{l}^{\alpha}=\mathcal{R}^{\dagger}l^{\alpha}\mathcal{R}.
\end{equation}
Here, $\mathcal{R}$ is the charge conjugation matrix 
\begin{equation}
\mathcal{R}=
\begin{pmatrix}
0 & 1 & 0 \\
1 & 0 & 0 \\
0 & 0 & 1
\end{pmatrix},
\end{equation}
with following properties:     
\begin{equation}
\mathcal{R}^{\dagger}=\mathcal{R}^t=\mathcal{R}^{-1}=\mathcal{R}.  
\end{equation}
The matrix $\mathcal{R}$ is diagonalized to yield  
\begin{equation}
-\kappa=
\begin{pmatrix}
-1 & 0 & 0\\
0 & 1 & 0 \\
0 & 0 & 1
\end{pmatrix},
\label{explicitk}
\end{equation}
and  the ``hermitian'' matrices that satisfy 
\begin{equation}
{\kappa^i}^{\dagger}=\kappa^i,~~{\kappa^{\alpha}}^{\dagger}=(\sigma^1)_{\beta}^{~~\alpha}\kappa^{\beta},
\label{hermiticykappaboth}
\end{equation}
are constructed by multiplying 
$\kappa$ to $l^i$ and $l^{\alpha}$: 
\begin{equation}
\kappa^i=2\kappa l^i,~~~~ \kappa^{\alpha}=2\kappa l^{\alpha}, 
\end{equation} 
or  
\begin{align}
&\kappa^1=-
\begin{pmatrix}
\sigma^2 &  0  \\
0 & 0 
\end{pmatrix},~~
\kappa^2=
\begin{pmatrix}
\sigma^1 & 0 \\
0 & 0 
\end{pmatrix},~~\kappa^3=
\begin{pmatrix}
1 & 0 \\
0 & 0 
\end{pmatrix},\nonumber\\
&\kappa^{\theta_1}=
\begin{pmatrix}
0 & \tau^{\theta_1} \\
-(\tau^{\theta_2})^t & 0 
\end{pmatrix},~~
\kappa^{\theta_2}=
\begin{pmatrix}
0 & -\tau^{\theta_2} \\
(\tau^{\theta_1})^t & 0 
\end{pmatrix}.
\end{align}

With the above preparation, we can realize the supersymmetric non-compact Hopf map (\ref{superhopfII}) as 
\begin{equation}
 \chi ~~\rightarrow~~ x^i=\chi^{\dagger}  \kappa^i\chi,~~\theta^{\alpha}=\chi^{\dagger} \kappa^{\alpha}\chi,
\label{SUSYHopfmapexpli}
\end{equation}
with  $\chi=(u,v,\eta)^t$ 
satisfying the ``normalization condition''   
\footnote{ Here, the symbols $\dagger$ and $*$ respectively
  represent the conventional hermitian conjugation and complex conjugation;  $\chi^{\dagger}=(u^*,v^*,\eta^*)$, and $(\eta^*)^*=\eta$, $(\eta_1\eta_2)^*=\eta_2^*\eta_1^*$. 
By the ``hermiticity'' of $\kappa^i$ and $\kappa^{\alpha}$ (\ref{hermiticykappaboth}), $x^i$ and $\theta^{\alpha}$ are ``real'' in the sense:  
${x^i}^*=x^i$ and ${\theta^{\alpha}}^*=(\sigma^1)_{\beta}^{~~\alpha}\theta^{\beta}.$ 
 Thus, $\theta=(\theta^1,\theta^2)^t$ is a $SO(2,1)$ Majorana-spinor. } 
\begin{equation}
\chi^{\dagger}\kappa\chi= u^*u-v^*v-\eta^*\eta= 1.
\label{normalizationpsi}
\end{equation}
The components of $x^i$ and $\theta^{\alpha}$ (\ref{SUSYHopfmapexpli}) are explicitly
\begin{align}
&x^1=iu^*v-iv^*u,~~~x^2=u^*v+v^*u,~~~x^3=u^*u+v^*v\nonumber\\
&\theta^1=u^*\eta-\eta^*v,~~~~~\theta^2=-v^*\eta+\eta^*u,
\end{align}
and denote coordinates on Euclidean $AdS^{2|2}$, since   
\begin{equation}
\eta_{ij}x^ix^j-\epsilon_{\alpha\beta}\theta^{\alpha}\theta^{\beta}
=-(\chi^{\dagger}\kappa\chi)^2=-1.
\end{equation}
Inverting (\ref{SUSYHopfmapexpli}) on the upper leaf of the $H^{2,0|2}$ ($x^3\ge 1$), $\chi$ is represented  as 
\begin{equation}
\chi=\frac{1}{\sqrt{2(1+x^3)}}
\begin{pmatrix}
({1+x^3})( 1-\frac{1}{4(1+x^3)}\theta\epsilon\theta )\\
({x^2-ix^1})(1+\frac{1}{4(1+x^3)}\theta\epsilon\theta)\\
(1+x^3)\theta^1+(x^2-ix^1)\theta^2
\end{pmatrix},
\label{explicitnonsusyhopfspinor}
\end{equation}
and the canonical connection of $U(1)$-fibre is 
\begin{equation}
A=-i\chi^{\dagger}\kappa d\chi=dx^i A_i +d\theta^{\alpha}A_{\alpha}, 
\end{equation}
where 
\begin{align}
&A_i=-\frac{1}{2}\epsilon_{ij3}\frac{x^j}{1+x^3}\biggl(1+\frac{2+x^3}{2(1+x^3)}\theta\epsilon\theta\biggr),\nonumber\\
&A_{\alpha}=-i\frac{1}{2}x^i (\theta \tau_i\epsilon)_{\alpha}.
\label{2superconneI}
\end{align}
The corresponding curvature  
\begin{equation}
F_{ij}=\partial_i A_j- \partial_j A_i,~~F_{i\alpha}=\partial_i A_{\alpha}-\partial_{\alpha}A_i,
~~F_{\alpha\beta}=\partial_{\alpha}A_{\beta}+\partial_{\beta}A_{\alpha} 
\end{equation}
is evaluated as 
\begin{align}
&F_{ij}=   -\frac{1}{2}\epsilon_{ijk}x^k(1+\frac{3}{2}\theta\epsilon\theta), \nonumber\\
&F_{i\alpha}=-i\frac{1}{2}(\theta\tau^j \epsilon)_{\alpha}
(\eta_{ij}-3x_ix_j), \nonumber\\
&F_{\alpha\beta}=-i(\tau_i \epsilon)_{\alpha\beta}x^i (1+
\frac{3}{2}\theta\epsilon\theta).
\label{superfieldstren2}
\end{align}
The non-singular canonical connection on the lower leaf ($x^3\le -1$)  is given by 
\begin{align}
&A'_i=\frac{1}{2}\epsilon_{ij3}\frac{x^j}{1-x^3}\biggl(1+\frac{2-x^3}{2(1-x^3)}\theta\epsilon\theta\biggr),\nonumber\\
&A'_{\alpha}=A_{\alpha}=-i\frac{1}{2}x^i (\theta  \tau_i \epsilon)_{\alpha},
\label{2superconneII}
\end{align}
and the corresponding curvature is the same (\ref{superfieldstren2}) 
\begin{equation}
F'_{ij}=F_{ij},~~F'_{i\alpha}=F_{i\alpha},~~F'_{\alpha\beta}=F_{\alpha\beta}. 
\end{equation}
The present base manifold is a super-extension  of the two-leaf hyperboloid, and as in the bosonic case, the topological structure of the canonical connection on each leaf might be trivial.  


\end{document}